\documentclass[entropy,article,submit,pdftex,moreauthors]{Definitions/mdpi}
\preto{\abstractkeywords}{\nolinenumbers}

\usepackage[utf8]{inputenc}
\firstpage{1} 
\pubvolume{1}
\issuenum{1}
\articlenumber{0}
\pubyear{2023}
\copyrightyear{2023}
\datereceived{9 October 2023} 
\daterevised{2 November 2023} 
\dateaccepted{4 November 2023} 
\datepublished{ } 
\hreflink{https://doi.org/} 

\DeclareMathOperator{\Tr}{Tr}

\Title{Work Fluctuations in Ergotropic Heat Engines}

\TitleCitation{Work Fluctuations in Ergotropic Heat Engines}

\Author{Giovanni 
 Chesi $^{1,}$*\orcidA{}, Chiara Macchiavello $^{1,2}$\orcidB{} and Massimiliano Federico Sacchi $^{2, 
3}$\orcidC{}}

\AuthorNames{Giovanni Chesi, Chiara Macchiavello and Massimiliano Federico Sacchi}

\AuthorCitation{Chesi, 
 G.; Macchiavello, C.; Sacchi, M.F.}

\address{%
$^{1}$ \quad National Institute for Nuclear Physics, Sezione di Pavia, Via Agostino Bassi 6, 27100 Pavia, Italy; chiara@unipv.it\\
$^{2}$ \quad QUIT Group, Dipartimento di Fisica, Università degli Studi di Pavia, Via Agostino Bassi 6, 27100 Pavia, Italy\\ 
$^{3}$ \quad CNR-Istituto di Fotonica e Nanotecnologie, Piazza Leonardo da Vinci 32, 20133 
 Milano, Italy; msacchi@unipv.it 
} 

\corres{Correspondence: giovanni.chesi@pv.infn.it}

\abstract{We study the work fluctuations in ergotropic heat engines, namely
two-stroke quantum Otto engines where the work stroke is designed to
extract the ergotropy (the maximum amount of work by a cyclic unitary
evolution) from a couple of quantum systems at canonical equilibrium
at two different temperatures, whereas the heat stroke thermalizes
back the systems to their respective reservoirs. We provide an
exhaustive study for the case of two qutrits whose energy levels are
equally spaced at two different frequencies by deriving the complete
work statistics. By varying the values of temperatures and
frequencies, only three kinds of optimal unitary strokes are found:
the swap operator $U_1$, an idle swap $U_2$ (where one of the qutrits
is regarded as an effective qubit), and a non-trivial permutation of
energy eigenstates $U_3$, which indeed corresponds to the composition
of the two previous unitaries, namely $U_3=U_2 U_1$. While $U_1$ and
$U_2$ are Hermitian (and hence involutions), $U_3$ is not. This point
has an impact on the thermodynamic uncertainty relations (TURs), which
bound the signal-to-noise ratio of the extracted work in terms of the
entropy production. In fact, we show that all TURs derived from a
strong detailed fluctuation theorem are violated by the transformation
$U_3$.}


\begin{document}

\section{Introduction}
A quantum description of thermodynamic heat engines has lately become
necessary to consider physical systems at the mesoscale and nanoscale
\cite{benny, li, stoc3}, such as nanojunctions thermoelectrics
\cite{dubi}, quantum dots~\cite{jose}, and biological~\cite{gnes8,
  ritort} or chemical~\cite{rao} systems. The optimal transport theory
has also recently been embedded in a thermodynamic quantum framework
\cite{saito}. At the quantum level, the fluctuations of the
thermodynamic variables play a fundamental role, due to the discrete
spectral structure of quantum systems.

The probability distributions of a set of thermodynamic variables
$\{X_i\}$ (energy, work, heat, particles,...) are related to the entropy production $\Sigma$ through the
so-called fluctuation theorems, which in general can be expressed as
\cite{galla, jar, crooks, piecho, jar2, j97, th, andrie, stoc2, sini,
  ciro, camp, mer, frq, cth, han, domenico, vo, moh}
\begin{linenomath}
    \begin{equation} \label{fluctgen}
        \frac{p(\{X_i\},\Sigma)}{p_B(\{-X_i\},-\Sigma)} = e^{\Sigma}
    \end{equation}
\end{linenomath}
where $p_B$ refers to the backward process, i.e., to the time-reversed
process identified by $p$. For a self-contained derivation of
  Equation~(\ref{fluctgen}) and its meaning in our context see Appendix~\ref{appA1} and
  Equation~(\ref{exch}). 
  There, a thermodynamical cycle is described by a
  set of stochastic trajectories which correctly reproduce the mean
  values $\{ \langle X_i\rangle \}, \langle \Sigma \rangle $
  of all variables $\{X_i \}, \Sigma $  by an average over all possible
  trajectories. Through the relation in Equation~(\ref{fluctgen},) the symmetries of the
processes set relevant constraints on the statistics of the variables
$\{X_i\}$. Another class of relations that connects the statistical
properties of mesoscopic and nanoscopic systems to the entropy
production is given by the so-called thermodynamic uncertainty
relations (TURs)~\cite{saito, vo, bar, ging, saito2, sam, domenico2,
  max, max2, landi, hase, pros, franci}. It has been shown that there
is a strong connection between fluctuation theorems and TURs,
i.e., every fluctuation theorem implies a specific TUR~\cite{sam}. Note
that the converse does not hold: it was recently found in
Ref.~\cite{domenico2} a TUR that does not stem from any fluctuation
theorem.

\begin{figure}[H]
    \includegraphics[width=6 cm]{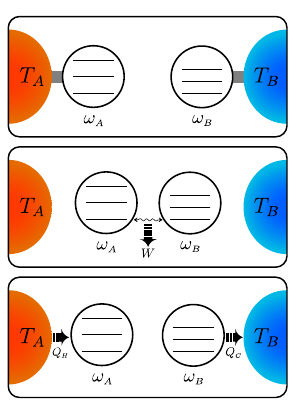}
    \caption{Scheme of a quantum thermodynamic engine based on the two-stroke Otto cycle with two qutrits as working fluid. In the first stage, the qutrits $A$ and $B$ with frequency $\omega_A$ and $\omega_B$ are at thermal equilibrium with the corresponding baths at temperature $T_A$ and $T_B$, respectively, with $T_A > T_B$. In the second stage, the two systems are isolated and allowed to interact through a unitary evolution extracting work $W$. Finally, in the last stage, the systems $A$ and $B$ are allowed to relax to the corresponding thermal baths, implying that $A$ absorbs the heat $Q_H$ and $B$ releases the heat $Q_C$, thus restoring the initial condition.}
    \label{otto}
\end{figure}

Thermodynamic engines that admit a straightforward quantum
description are the ones based on the Otto cycle~\cite{moh, max, max2,
  kosloff2, kosloff, grama, nori, thomas, abah, jukka, peterson, moli,
  bruno, kuz} since the work and heat exchanged are unambiguously
identified by their respective distinct strokes. The case considered
in this paper, namely a two-stroke Otto cycle, is outlined in
Figure~\ref{otto}, where the working fluid is represented by two
qutrits.
\vspace{-6pt}
In the case of an engine based on a two-stroke Otto cycle, the full probability distribution of work and heat has been retrieved for two qudits~\cite{max} and for two bosonic modes~\cite{max2} as working fluids, where the transformation for the work extraction is the unitary partial-swap interaction. The two-stroke Otto engine is particularly interesting with respect to its well-known four-stroke version because it allows the extraction of the maximum amount of work in the adiabatic step of the cycle by a single unitary operation, the so-called ergotropy~\cite{allah, allah2, ando1, jack, giova, horo, ando2}. Note that the extraction of the ergotropy necessarily also depends on the transformation that couples the systems. We show here that if the systems are qudits with dimensions larger than two, unitary evolutions different from the swap interaction can increase the extracted work.

We define a procedure for determining the unitary interaction that provides the maximum work from two multilevel systems $A$ and $B$ for a given choice of the relevant parameters, i.e., the frequency gaps $\omega_A$ and $\omega_B$ of the qudits and the temperatures $T_A$ and $T_B$ of the reservoirs. Then we take the specific case of a working fluid described by two qutrits and classify all the transformations that extract the ergotropy. Specifically, we find three different kinds of optimal unitary strokes: the swap operator $U_1$, an idle swap $U_2$ (where one of the qutrits is regarded as an effective qubit), and a non-trivial permutation $U_3$ given by a composition of the two previous unitaries, namely $U_3=U_2 U_1$. Each transformation extracts the ergotropy from a different regime defined by the frequency gaps $\omega_A$ and $\omega_B$ of the two qutrits and by the temperatures $T_A$ and $T_B$ of the baths. By deriving the characteristic function of work and heat, we evaluate the work statistics and the entropy production for every case. Note that a complete description of a quantum ergotropic heat engine and of the procedure for determining the work statistics is detailed in Appendix~\ref{appA1}. Then, we focus on the trade-off between ergotropy extraction and relative fluctuations $\text{var}(W)/\langle W \rangle^2$, i.e., the inverse of the signal-to-noise ratio (SNR). The evaluation of the fluctuations allows us to establish the relation between the variance of the work and the mean entropy production in terms of the TURs.
A standard reference TUR bounds the fluctuations with the inverse of the entropy production as follows~\cite{bar}
\begin{linenomath}
    \begin{equation} \label{fluct}
        \frac{\text{var}(W)}{\langle W \rangle^2} \geq \frac{2}{\langle\Sigma\rangle}.
    \end{equation}
\end{linenomath}
We show that all three ergotropic transformations violate this TUR. Moreover, $U_3$ is proved to beat all the TURs derived by the strong fluctuation theorem where the forward and backward processes in Equation~(\ref{fluctgen}) are related by the same condition $p_B(\{X_i\}, \Sigma) = p(\{X_i\},\Sigma)$.

This paper is structured as follows. In Section~\ref{mm}, we define the procedure for determining the transformations extracting the ergotropy in the case where the working fluid is described by two qudits with generic dimensions $d_A$ and $d_B$. Then, in Section~\ref{r}, we apply our procedure to the case of two qutrits. In particular, in Section~\ref{r1}, we classify all the transformations extracting the ergotropy and their properties. In Section~\ref{r2}, we evaluate the maximum work extracted by each transformation in terms of the frequency gaps and the temperatures. In Section~\ref{r3}, we study the mean entropy production related to each interaction. In Section~\ref{r4}, we derive the work distributions. Finally, in Section~\ref{r5}, we find the relative fluctuations of work, compare the corresponding SNR to the bounds provided by the most relevant TURs, and discuss the assumptions required for these TURs to hold. In Section~\ref{c}, we draw our conclusions.

\section{Materials and Methods} \label{mm}
\textls[-15]{In this work, we fix the Planck and Boltzmann constants to natural units, i.e., \mbox{$\hbar=k_B=1$.}}
We consider two qudits $A$ and $B$ in a product of Gibbs states, i.e.,
\begin{linenomath}
\begin{equation} \label{state}
\rho_{0} = \frac{e^{-\beta_AH_A}}{Z_A}\otimes\frac{e^{-\beta_BH_B}}{Z_B}
\end{equation}
\end{linenomath}
where $H_X = \omega_X\sum_{n=0}^{d_X-1} n|n\rangle\langle n|$ is the
Hamiltonian of the system $X=A,B$, each one with equally-spaced energy
levels, and $Z_X = \text{Tr}[e^{-\beta_XH_X}]$ denotes the
corresponding partition function, and $\beta_X = T_X^{-1}$ the inverse
temperature. The number states $|n\rangle$ in the expansion of the
Hamiltonians are eigenstates of the occupation number $n_X \equiv
H_X/\omega_X$. Without loss of generality, we fix $T_A > T_B$.

We 
 use the state in Equation~(\ref{state}) as the input to a two-stroke
Otto engine. As depicted in Figure~\ref{otto}, the process starts with
the two qudits in thermal equilibrium with their baths, at temperature
$T_A$ and $T_B$. Afterwards, the two qudits are isolated from their
baths and we make them interact through a unitary evolution in order
to extract the ergotropy. The procedure for the ergotropy extraction
will be detailed in the following. Once the work has been extracted
through the interaction, the two qudits are
  decoupled from each other and then reset to their equilibrium states,
namely as in Equation~(\ref{state}), by re-connecting them to their thermal
baths via a weak-coupling and energy-preserving
  interaction. In this way, no work contribution comes from the on-off
  interaction of the systems with the reservoirs~\cite{jukka,gabri,moli}. The sequential repetition of this process
  leads to our two-stroke cyclic engine.

We 
 fix the convention
of positive work for the extraction from the system and positive heat
for the absorption from the reservoirs. Then, in each cycle the
average energy change in system $A$ due to the unitary stroke
corresponds to the average heat released by the hot reservoir $A$,
namely $\langle Q_H \rangle = - \langle \Delta E_A\rangle$. Similarly,
for the cold reservoir, $\langle Q_C \rangle = - \langle \Delta
E_B\rangle$, and, for the first law of thermodynamics, the average
work is given by $\langle W \rangle= \langle Q_H \rangle + \langle Q_C
\rangle = -\langle \Delta E_A \rangle-\langle\Delta E_B\rangle$.
Correspondingly, the average entropy production reads $\langle \Sigma
\rangle = -\beta_A\langle Q_H\rangle-\beta_B\langle Q_C \rangle =
(\beta_A - \beta_B)\langle \Delta E_A\rangle - \beta_B\langle
W\rangle$. Our goal is the investigation of an ergotropic heat engine
based on the two-qudit system described above, i.e., an engine
extracting the maximum work by exploiting the difference in frequency
and temperature between the systems $A$ and $B$. In other words, we
are looking for the unitary transformations $U$ mapping the input
$\rho_0$ into a state $\rho = U\rho_0U^{\dagger}$ such that the
average extracted work is maximized, i.e.,
\begin{linenomath}
\begin{equation} \label{maxW}
\langle W \rangle = \max_{U}\{\text{Tr}[\rho_0 H] - \text{Tr}[\rho H]\}
\end{equation}
\end{linenomath}
where $H=H_A\otimes \mathbb{I}_B +\mathbb{I}_A \otimes H_B$ is
the Hamiltonian of the system. The evolution that extracts the
ergotropy was identified in Ref.~\cite{allah} as the one minimizing the final energy $\text{Tr}[\rho H]$. In the present case, where the initial state $\rho_0$ has no coherence, namely, it is diagonal in the energy basis, the ergotropic evolution is the transformation that permutes the eigenstates of the input state so that the magnitude order of the energy levels is reversed with respect to the corresponding occupation fractions. More explicitly, if we take the occupation fractions of the system $e^{-(n\beta_A \omega_A+m\beta_B\omega_B)}/(Z_AZ_B)$ in descending order, the transformation permutes the related eigenstates to set the corresponding energy levels in ascending order. If the input state already displays this configuration, then the state is called passive and no transformation can extract work. In summary, since unitary transformations preserve the spectrum, the ergotropy is extracted by reversing all possible population inversion with respect to the energy levels. In the following, we provide a re-visited analysis of the first-level maximization strategy developed in Ref.~\cite{allah2}.

The 
 procedure of ergotropy extraction can be formalized in a compact way for two subsystems $A$ and $B$ of dimension $d_A$ and $d_B$ as follows. We consider two different permutations $P_E$ and $P_{\rho}$ of the energy eigenstates with respect to their lexicographic order. The permutation $P_E$ sorts them so that the corresponding eigenvalues are set in ascending order,~i.e.,
\begin{linenomath}
    \begin{equation}
        \begin{aligned}
            P_E H P_E^{\dagger} = P_E \left(\sum_{j=0}^{d_A-1}\sum_{k=0}^{d_B-1}(E_j + E_k)|jk\rangle\langle jk|\right) P_E^{\dagger} = \sum_{l=0}^{d_A d_B - 1} \tilde{E}_l |l\rangle\langle l | \equiv H^{\uparrow}
        \end{aligned}
    \end{equation}
\end{linenomath}
where the vector of eigenvalues $\pmb{\tilde{E}} = \{\tilde{E}_l\}_{l=0}^{d_A d_B - 1}$ satisfies $\tilde{E}_l < \tilde{E}_{l+1} \,\, \forall\,l\in[0,d_A d_B - 1)$. Similarly, the permutation $P_{\rho}$ rearranges the occupation numbers of the initial state in descending order, namely,
\begin{linenomath}
    \begin{equation}
        \begin{aligned}
            P_{\rho} \rho_0 P_{\rho}^{\dagger} = P_{\rho} \left(\sum_{l=0}^{d_A d_B - 1} r_l |l\rangle\langle l |\right)P_{\rho}^{\dagger} = \sum_{l=0}^{d_A d_B - 1} \tilde{r}_l |l\rangle\langle l | \equiv \rho_0^{\downarrow}
        \end{aligned}
    \end{equation}
\end{linenomath}
and $\pmb{\tilde{r}} = \{\tilde{r}_l\}_{l=0}^{d_A d_B - 1}$ is such that $\tilde{r}_{l+1} < \tilde{r}_l \,\, \forall\,l\in[0,d_A d_B - 1)$. Then, we can straightforwardly find the transformation that minimizes the final energy from
\begin{linenomath}
    \begin{equation}
        \begin{aligned}
            \text{Tr}[\rho H] &= \text{Tr}[U\rho_0 U^{\dagger} H] = \text{Tr}[\rho_0^{\downarrow} H^{\uparrow}] = \text{Tr}[P_{\rho} \rho_0 P_{\rho}^{\dagger} P_E H P_E^{\dagger}] \\
            &= \text{Tr}[P_E^{\dagger}P_{\rho} \rho_0 P_{\rho}^{\dagger} P_E H]
        \end{aligned}
    \end{equation}
\end{linenomath}
implying that the ergotropic transformation can be expressed as
\begin{linenomath}
    \begin{equation}
        \begin{aligned}
            U = P_E^{\dagger}P_{\rho}.
        \end{aligned}
    \end{equation}
\end{linenomath}
\\
For instance, take two qubits in a Gibbs state
\begin{linenomath}
\begin{equation} \label{stateex}
\rho_{0} = \frac{1}{Z_AZ_B}\sum_{n,m=0}^1e^{-n\beta_a\omega_A-m\beta_B\omega_B}|nm\rangle\langle nm|.
\end{equation}
\end{linenomath}
Then, the energies pertaining to the levels $|10\rangle\langle 10|$ and $|01\rangle\langle 01|$ are $\omega_A$ and $\omega_B$, respectively, while the related occupation fractions are $Z_A^{-1}Z_B^{-1}e^{-\beta_A\omega_A}$ and $Z_A^{-1}Z_B^{-1}e^{-\beta_B\omega_B}$. If we have $\omega_A>\omega_B$ and $\beta_A\omega_A < \beta_B\omega_B$ or the symmetric case where both the order relations are reversed, the transformation that swaps $|10\rangle\langle 10|$ with $|01\rangle\langle 01|$, namely $U = U^{\dagger} = |00\rangle\langle 00|+|11\rangle\langle 11| + |01\rangle\langle 10| + |01\rangle\langle 10|$, extracts the ergotropy. This result appears immediately if we consider the permutation matrices $P_E$ and $P_{\rho}$, which in this case read
\begin{linenomath}
    \begin{equation}
        \begin{aligned}
            &P_E = \theta(\omega_A-\omega_B)\mathbb{I} + \theta(\omega_B-\omega_A)U \\
            &P_{\rho} = \theta(\beta_A\omega_A-\beta_B\omega_B)\mathbb{I} + \theta(\beta_B\omega_B-\beta_A\omega_A)U
        \end{aligned}
    \end{equation}
\end{linenomath}
where $\theta(x)$ is the Heaviside function. The operator $P_E^{\dagger}P_{\rho}$ promptly identifies the ergotropic transformations and the corresponding ergotropic regimes, since
\begin{linenomath}
    \begin{equation}
    \begin{aligned}
        P_E^{\dagger}P_{\rho} =& [\theta(\omega_A-\omega_B)\theta(\beta_A\omega_A-\beta_B\omega_B)+\theta(\omega_B-\omega_A)\theta(\beta_B\omega_B-\beta_A\omega_A)]\mathbb{I}\, + \\
        & [\theta(\omega_A-\omega_B)\theta(\beta_B\omega_B-\beta_A\omega_A) + \theta(\omega_B-\omega_A)\theta(\beta_A\omega_A-\beta_B\omega_B)]U.
    \end{aligned}
    \end{equation}
\end{linenomath}
\\
This simple example shows how the extraction of the ergotropy is entirely determined by the order relations between the parameters. In particular, the initial state of an equally-spaced two-qudit engine is described for any dimension of the qudits by a first partial order over the frequencies $\omega$ and a second one over the products $\beta\omega$. These order relations identify four basic partially ordered sets (posets). In the two-qubit example, the ergotropy can only be extracted if the initial state belongs to $\Omega \equiv \{\omega_A > \omega_B\ \wedge \beta_A\omega_A < \beta_B\omega_B\}$ or $\Bar{\Omega}$, where the bar denotes the same poset with $A$ and $B$ switched. The states belonging to the remaining two sets are passive.

The 
 description in terms of posets becomes more complex in higher dimensions. For a state as in Equation~(\ref{state}), the ordering procedure for the ergotropy extraction needs to establish if $k\omega_A > j\omega_B$ and if $k\beta_A\omega_A > j\beta_B\omega_B$ for every pair of natural numbers $k \in [0,d_A)$ and $j \in [0,d_B)$.

Even 
 if the simplest non-trivial case would be a system made of a
qubit and a qutrit, here, as mentioned above, we consider a two-qutrit
system, so that we can use the results for the two-stroke swap Otto
engine with two qudits with equal dimensions studied in Ref.~\cite{max}
as a benchmark. In this scenario, each of the four basic posets mentioned above is further partitioned in four subsets, defined by the order relations $0< y_{X_1} < y_{X_2}/2$ and $y_{X_2}/2 < y_{X_1} < y_{X_2}$, with $y=\omega$ or $\beta\omega$ and $X_1 \neq X_2$ may be $A$ or $B$. The total number of posets determining the regimes for the ergotropy extraction is then sixteen. We expect some of them to be passive regimes, i.e., the input state defined by those parameters is passive. As for the others, we will show that a specific transformation can extract the ergotropy from different regimes, as we noted for the two-qubit case with the swap in the regimes $\Omega$ and $\bar{\Omega}$. 

\section{Results} \label{r}

\subsection{Ergotropic Transformations} \label{r1}
As mentioned above, we can jointly classify all the ergotropic transformations $U$ and the corresponding ergotropic regimes by inspecting the permutations $P_E$ and $P_{\rho}$.

\begin{figure}[H]
\includegraphics[width=13.5 cm]{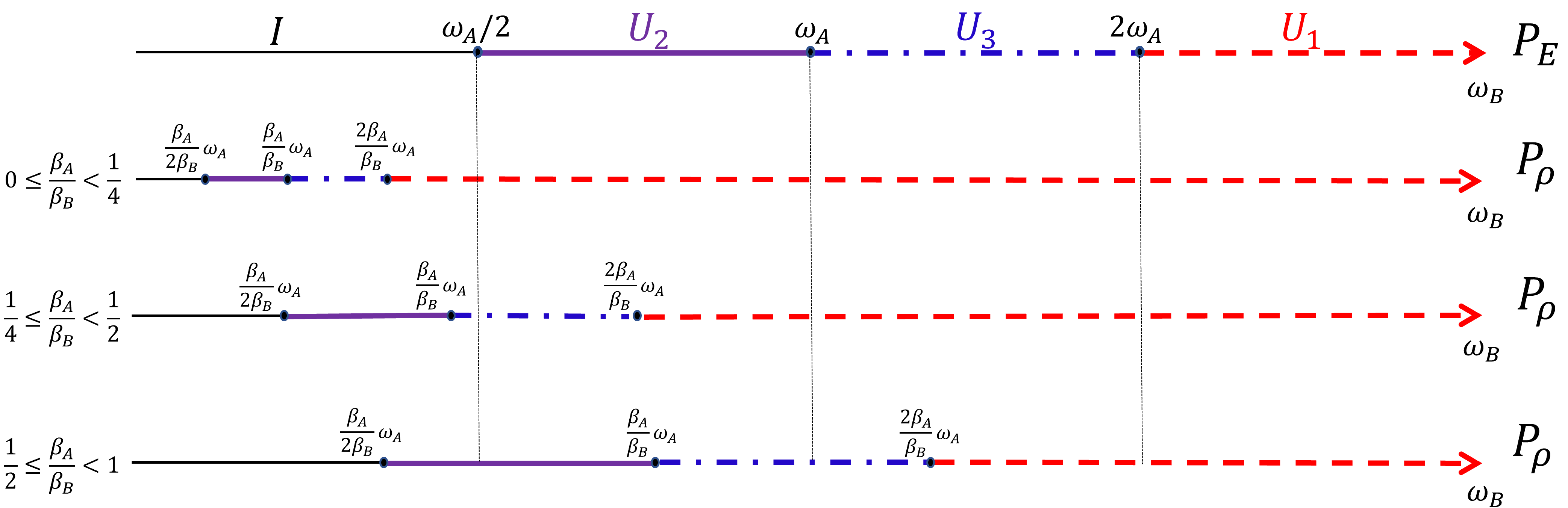}
\caption{Scheme of the transformations realizing the permutations $P_E$ and $P_{\rho}$ in the different regimes identified by $\omega$ in the former case and by $\beta\omega$ in the latter. We show these regimes by fixing $\omega_A$ and the three inequivalent cases for the temperature ratio $\beta_A/\beta_B$ and studying $P_E$ and $P_{\rho}$ for increasing $\omega_B$. As $\omega_B$ increases, we find that both the permutations are given by the identity (black thin line), $U_2$ (purple thick line), $U_3$ (blue dotdashed line), and the swap $U_1$ (red dashed line).    \label{permap}}
\end{figure}

In 
 the two-qutrit case we have four posets identified by $\omega_A$ and $\omega_B$ for $P_E$, and four identified by $\beta_A\omega_A$ and $\beta_B\omega_B$ for $P_{\rho}$. We find different permutations $P_E$ and $P_{\rho}$ for each of the corresponding four posets, i.e., four distinct transformations. We show them associated with the corresponding poset in Figure~\ref{permap}. Note that, for what concerns $P_{\rho}$, we have to distinguish three inequivalent cases identified by the relative position of points on the $\omega_B$ axis according to the value of the ratio $\beta_A/\beta_B$.

In 
 summary, $P_E$ and $P_{\rho}$ are simply the identity $I$ (i.e., no reordering is needed) for $\omega_A < 2\omega_B$ and $\beta_A\omega_A < 2\beta_B\omega_B$, respectively. For $\omega_B > 2\omega_A$ and $\beta_B\omega_B > 2\beta_A\omega_A$, both $P_E$ and $P_{\rho}$ are given by the swap $U_1$, namely
\begin{linenomath}
\begin{equation} \label{u1}
\begingroup\makeatletter\def\f@size{9.5}\check@mathfonts
\def\maketag@@@#1{\hbox{\m@th\normalsize\normalfont#1}}%
\begin{aligned}
U_1 = U_1^{\dagger} =& |00\rangle\langle 00| + |11\rangle\langle 11| + |22\rangle\langle 22| + |01\rangle\langle 10| + |10\rangle\langle 01| + |02\rangle\langle 20| + |20\rangle\langle 02| + \\
    &|12\rangle\langle 21| + |21\rangle\langle 12|,
\end{aligned}
\endgroup
\end{equation}
\end{linenomath}
or, equivalently, $U_1 = (24)(37)(68)$, using the cycle notation and the lexicographic ordering where the elements of the cycles are related to the kets as $|nm\rangle\rightarrow 3n+m+1$.

For $\omega_B \in [\omega_A/2,\omega_A]$ and $\beta_B\omega_B \in [\beta_A\omega_A/2, \beta_A\omega_A]$, both $P_E$ and $P_{\rho}$ are given by
\begin{linenomath}
\begin{equation} \label{u2s}
\begingroup\makeatletter\def\f@size{9.5}\check@mathfonts
\def\maketag@@@#1{\hbox{\m@th\normalsize\normalfont#1}}%
\begin{aligned}
U_2 = U_2^{\dagger} =& |00\rangle\langle 00| + |11\rangle\langle 11| + |22\rangle\langle 22| + |01\rangle\langle 01| + |21\rangle\langle 21| + |10\rangle\langle 02| + |02\rangle\langle 10| + \\
    &|20\rangle\langle 12| + |12\rangle\langle 20| = (34)(67).
\end{aligned}
\endgroup
\end{equation}
\end{linenomath}
Finally, for $\omega_B \in [\omega_A, 2\omega_A]$ and $\beta_B\omega_B \in [\beta_A\omega_A, 2\beta_A\omega_A]$, both $P_E$ and $P_{\rho}$ are given by
\begin{linenomath}
    \begin{equation} \label{u3}
\begingroup\makeatletter\def\f@size{9.5}\check@mathfonts
\def\maketag@@@#1{\hbox{\m@th\normalsize\normalfont#1}}%
        \begin{aligned}
            U_3 =& |00\rangle\langle 00| + |11\rangle\langle 11| + |22\rangle\langle 22| + |01\rangle\langle 10| + |10\rangle\langle 20| + |20\rangle\langle 21| + |21\rangle\langle 12| + \\
            &|12\rangle\langle 02| + |02\rangle\langle 01| = (236874).
        \end{aligned}
        \endgroup        
    \end{equation}
\end{linenomath}
We notice that $U_2$ and $U_3$ are not invariant under swap symmetry. In particular, $\tilde{U}_2 \equiv U_1U_2U_1 = U_1U_3$ reads
\begin{linenomath}
\begin{equation} \label{u2}
\begingroup\makeatletter\def\f@size{9.5}\check@mathfonts
\def\maketag@@@#1{\hbox{\m@th\normalsize\normalfont#1}}%
\begin{aligned}
\tilde{U}_2 =\tilde{U}_2^{\dagger} =& |00\rangle\langle 00| + |11\rangle\langle 11| + |22\rangle\langle 22| + |10\rangle\langle 10| + |12\rangle\langle 12| + |01\rangle\langle 20| + |20\rangle\langle 01| + \\
    &|02\rangle\langle 21| + |21\rangle\langle 02| = (27)(38),
\end{aligned}
\endgroup
\end{equation}
\end{linenomath}
while 
\begin{linenomath}
    \begin{equation} \label{u3s}
        \tilde{U}_3 \equiv U_1 U_3 U_1 = U_3^{-1} = U_3^{\dagger}.
    \end{equation}
\end{linenomath}
The unitary operators $U_1$, $U_2$, and $\tilde{U}_2$ are also Hermitian and hence self-inverse. Notice also~that
\begin{linenomath}
    \begin{equation}
        U_3 = U_1 \tilde{U}_2 = U_2 U_1,
    \end{equation}
\end{linenomath}
and, similarly, $\tilde{U}_3 = U_1 U_2 = \tilde{U}_2 U_1$.

\begin{figure}[H]
\includegraphics[width=10.5 cm]{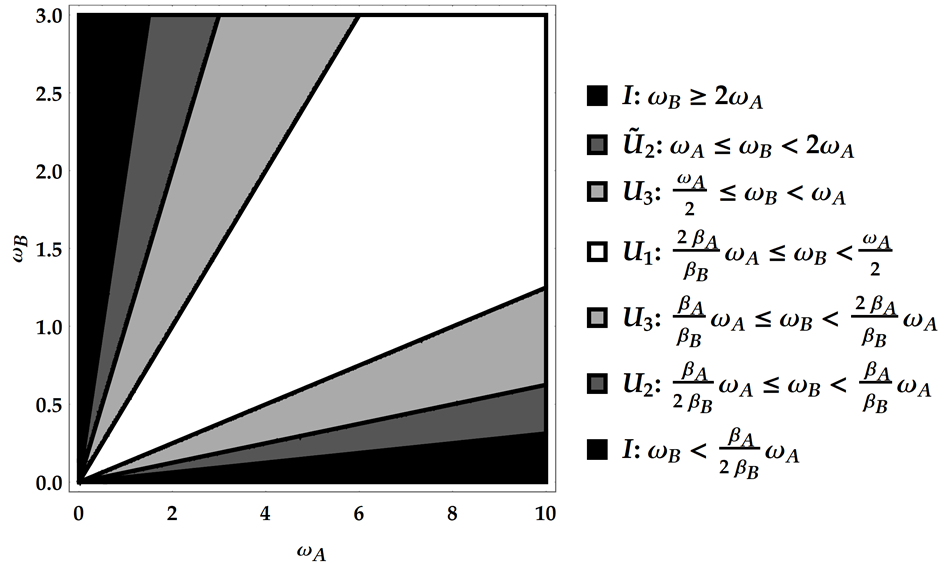}
\caption{First case: $0 < \beta_A/\beta_B < 1/4$. Here, specifically, $\beta_A/\beta_B = 1/16$. \label{case1}}
\end{figure}

\begin{figure}[H]
\includegraphics[width=10.5 cm]{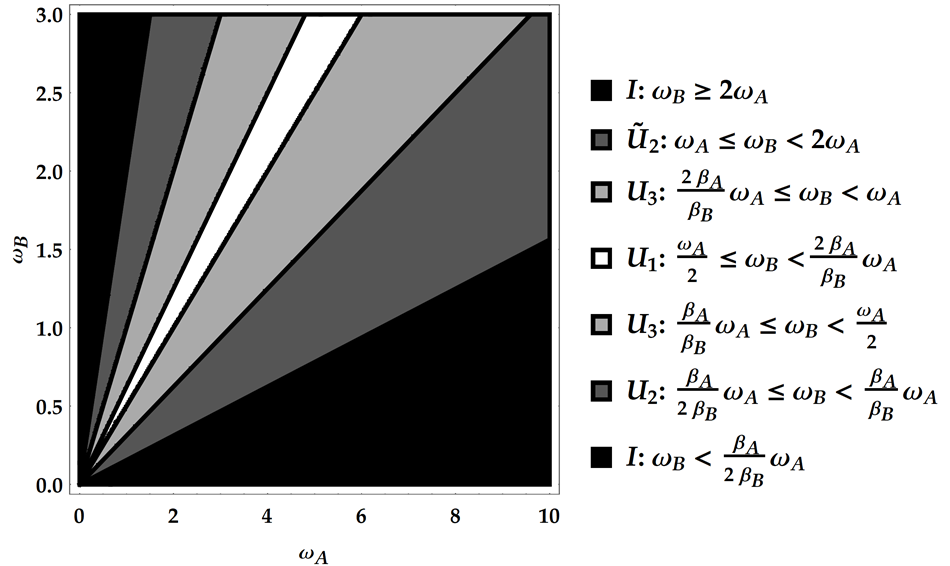}
\caption{Second case: $1/4 < \beta_A/\beta_B < 1/2$. Here, specifically, $\beta_A/\beta_B = 5/16$. \label{case2}}
\end{figure}

\begin{figure}[H]
\includegraphics[width=10.5 cm]{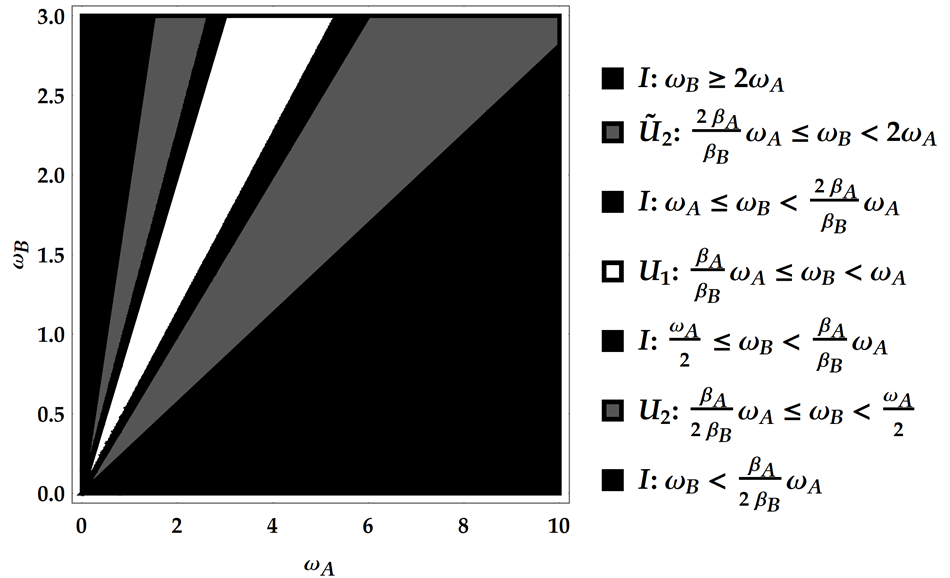}
\caption{Third case: $1/2 < \beta_A/\beta_B < 1$. Here, specifically, $\beta_A/\beta_B = 9/16$. \label{case3}}
\end{figure}

The 
 product $P_E^{\dagger} P_{\rho}$ together with the composition rules for $U_1$, $U_2$ and $U_3$ explored above allows to find the ergotropic transformations for each ergotropic regime identified by combining an $\omega$ poset with a $\beta\omega$ poset. In particular, we remark that the ergotropic transformations resulting from the product $P_E^{\dagger} P_{\rho}$ must be again $U_1$, $U_2$, $\tilde{U}_2$ and $U_3$. There are five overall, considering the identity too, which pertains to initial passive states. We provide a direct visualization of the landscape of ergotropic transformations in Figures~\ref{case1}--\ref{case3}. Having set $\beta_A < \beta_B$, each figure is linked to a different regime for the ratio $\beta_A/\beta_B$. As outlined in Figure~\ref{permap}, we can identify three distinct ranges of $\beta_A/\beta_B$ with two critical values, namely $1/4$ and $1/2$. 
For each case, we show the ergotropic transformation related to each poset. In particular, we set $\beta_A/\beta_B = 1/16$ in Figure~\ref{case1}, $\beta_A/\beta_B = 5/16$ in Figure~\ref{case2} and $\beta_A/\beta_B = 9/16$ in Figure~\ref{case3}. Firstly, we observe that in the first two cases, all the transformations found above appear (except $\tilde{U}_3$, which pertains to the regime $T_A < T_B$). In the case of Figure~\ref{case3}, $U_3$ is never present and the number of passive regimes becomes four. Notice that in the region $0 < \beta_A/\beta_B < 1/4$, it is possible to take the limits $\beta_A \rightarrow 0$ and $\beta_B \rightarrow \infty$. In this case, one of the passive regimes disappears and most of the parameter region is dominated by the swap. On the contrary, approaching the critical point $\beta_A/\beta_B = 1/4$ we see that the region where the swap extracts the ergotropy shrinks until it vanishes at the critical point. In the second case, in Figure~\ref{case2}, the swap plays again a role, but the passive regimes grow as well until, at the critical point $\beta_A/\beta_B = 1/2$, the ergotropic region of $U_3$ vanishes and is replaced for $\beta_A/\beta_B > 1/2$ by passive regimes. Of course, at $\beta_A/\beta_B = 1$, the whole frequency subset is~passive.

Let us inspect more in detail the non-trivial ergotropic transformations $U_1$, $U_2$, $\tilde{U}_2$ and $U_3$.
The swap $U_1$ clearly commutes with the total number operator, namely
\begin{linenomath}
    \begin{equation} \label{comm1}
        [U_1, n_A\otimes \mathbb{I}_B + \mathbb{I}_A\otimes n_B] = 0.
    \end{equation}
\end{linenomath}
On the other hand, the evolutions $U_2$ and $\tilde{U}_2$ act asymmetrically on the two systems, 
since they perform a permutation of the frequency levels of $\rho_0$ as if the system identified by the smallest frequency gap ($B$ when the ergotropy is extracted by $U_2$ and $A$ when it is extracted by $\tilde{U}_2$) were a two-level system, being its intermediate level $|1\rangle$ left unaffected. Thus, we name $U_2$ as \textit{idle swap}. In fact, for this asymmetry, we have $U_2 \neq \tilde{U}_2$.

Differently from $U_1$, the idle swaps $U_2$ and $\tilde{U}_2$ enjoy the conservation laws 

\begin{linenomath}
    \begin{equation} \label{comm2}
        \begin{aligned}
            &[U_2, 2n_A\otimes \mathbb{I}_B + \mathbb{I}_A\otimes n_B] = 0, \\
            &[\tilde{U}_2, n_A\otimes \mathbb{I}_B + \mathbb{I}_A\otimes 2n_B] = 0.
        \end{aligned}
    \end{equation}
\end{linenomath}
As for $U_3 = U_2 U_1$, being the composition of the standard and the idle swap, we name it \textit{double swap}. We noticed above that $U_3$ is not Hermitian. Indeed, one finds out that the double swap has multiplicative order six, namely $U_3^{6}=\mathbb{I}$, as it can be inferred from the cycle notation in Equation~(\ref{u3}). Furthermore, the double swap does not commute with any linear combination of $n_A$ and $n_B$. In Appendix~\ref{appA1}, we prove that, if the transformation commutes with a linear combination of $H_A$ and $H_B$, then all work and heat moments are proportional to each other, and hence, the mean entropy production is proportional to the mean extracted work, as we will explicitly show for $U_1$, $U_2$ and $\tilde{U}_2$ in the next sections.

\unskip

\subsection{Ergotropy} \label{r2}
\begin{figure}[H]
    \includegraphics[width=8 cm]{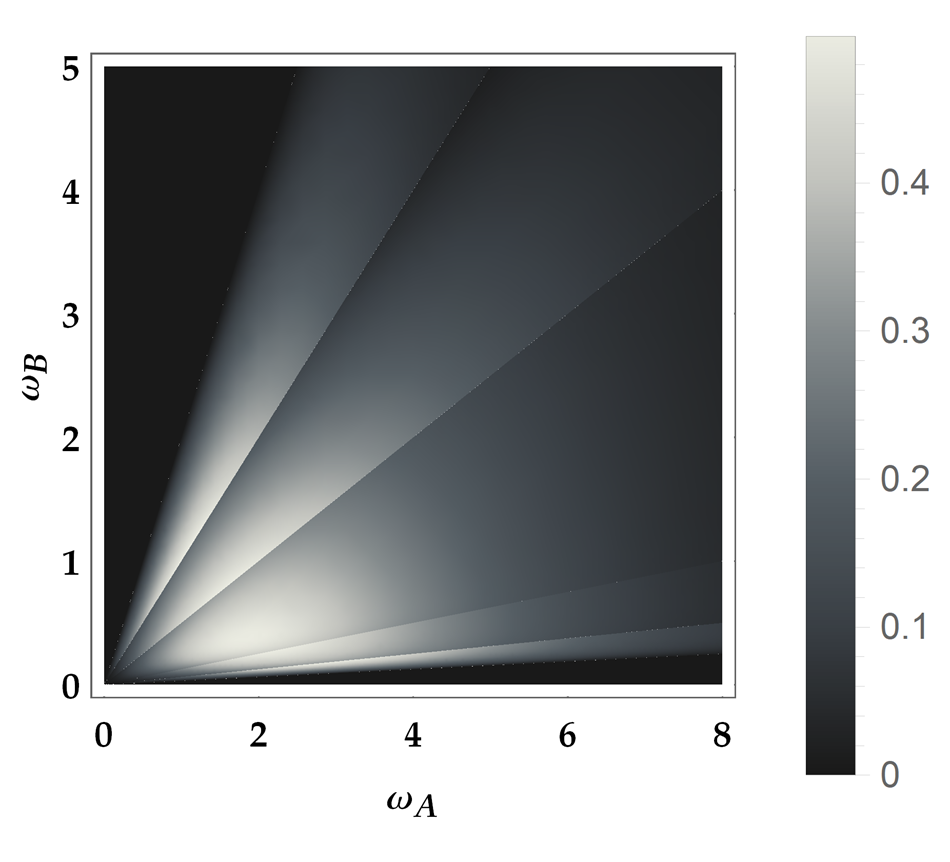}
    \caption{Ergotropy extraction in the case $0 < \beta_A/\beta_B < 1/4$. Here, $\beta_A/\beta_B = 1/16$ and $\beta_B = 10$.}
    \label{wereg1}
\end{figure}

\begin{figure}[H]
    \includegraphics[width=10.5 cm]{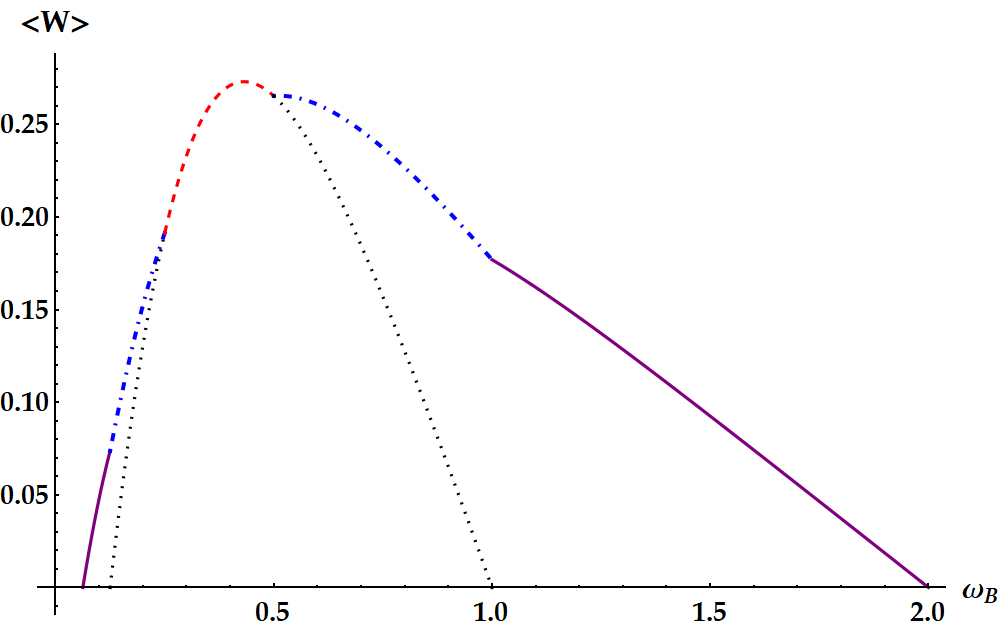}
    \caption{Ergotropy $\langle W \rangle$ as a function of $\omega_B$ in the case $\beta_A/\beta_B=1/8$, with $\omega_A = 1$, $\beta_A = 0.5$, $\beta_B = 4$. Purple solid lines: idle swaps $U_2$ ($\omega_B < 1/8$) and $\tilde{U}_2$ ($\omega_B > 1$). Blue dot-dashed line: double swap $U_3$. Red dashed line: standard swap $U_1$ inside the corresponding ergotropic regime. Black dotted line: standard swap for any $\omega_B$ such that the extracted work is positive.}
    \label{es2}
\end{figure}

\begin{figure}[H]
    \includegraphics[width=10.5 cm]{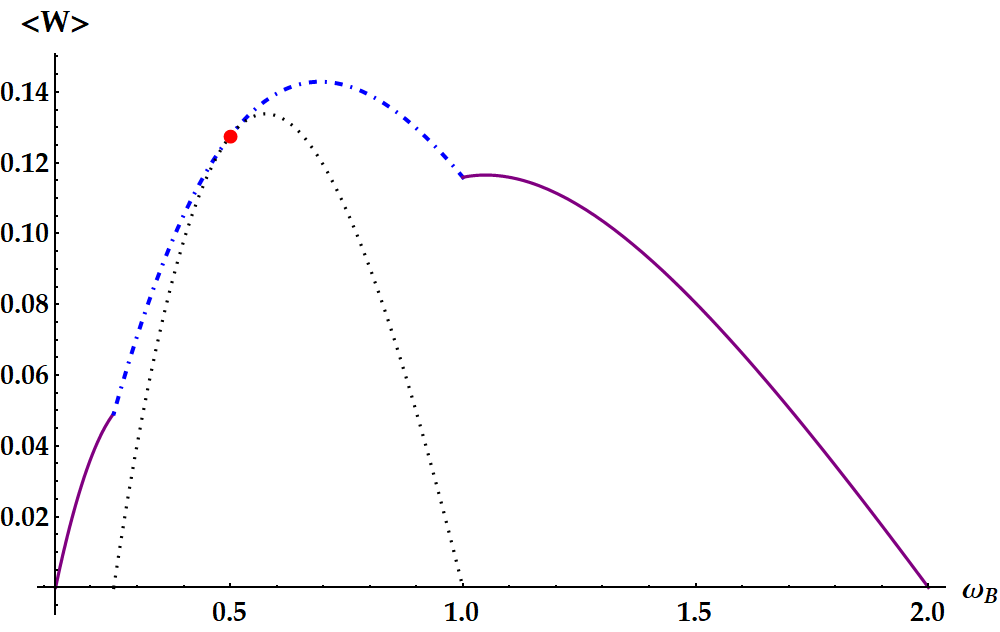}
    \caption{Ergotropy $\langle W \rangle$ as a function of $\omega_B$ at the critical point $\beta_A/\beta_B=1/4$, with $\omega_A = 1$, $\beta_A = 0.5$, $\beta_B = 2$. The red mark identifies the tangent point where the standard swap $U_1$ (dotted black line) and the double swap $U_3$ (blue dashed-dotted line) extracts the same amount of work at $\omega_B = \omega_A/2 = 0.5$. The purple solid curves identify the ergotropy extracted by $U_2$ ($1/8 < \omega_B < 1/4$) and $\tilde{U}_2$ ($1 < \omega_B < 2$).}
    \label{es3}
\end{figure}

\begin{figure}[H]
    \includegraphics[width=9.3 cm]{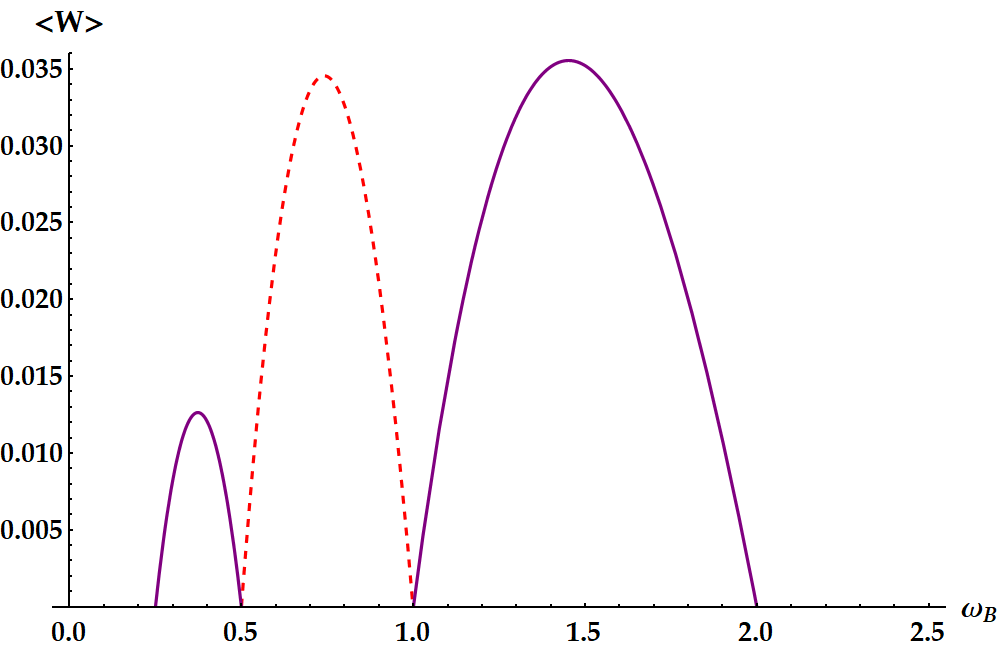}
    \caption{Ergotropy $\langle W \rangle$ as a function of $\omega_B$ in the critical case $\beta_A/\beta_B=1/2$, with $\omega_A = 1$, $\beta_A = 0.5$, $\beta_B = 1$. Dashed red line: standard swap $U_1$ for $\omega_B \in [\omega_A/2,\omega_A] = [1/2,1]$. Purple solid lines: idle swaps $U_2$ for $\omega_B \in [\beta_A\omega_A/2\beta_B,\omega_A/2] = [1/4,1/2]$ and $\tilde{U}_2$ for $\omega_B \in [\omega_A, 2\omega_A] = [1,2]$.}
    \label{es1}
\end{figure}

\begin{figure}[H]
    \includegraphics[width=9.3 cm]{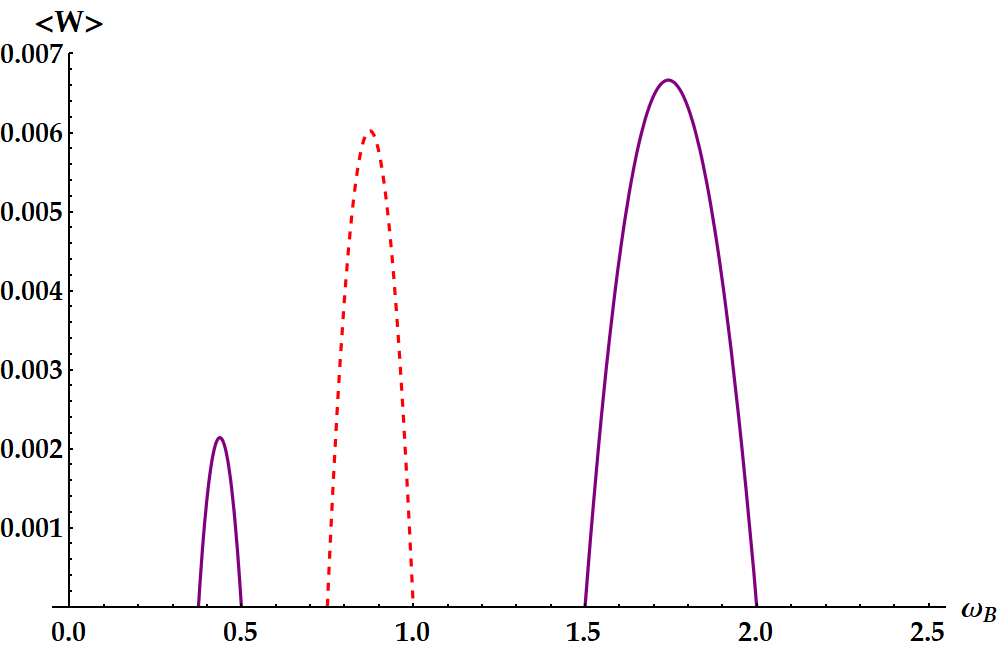}
    \caption{Ergotropy $\langle W \rangle$ as a function of $\omega_B$ in the case $\beta_A/\beta_B=3/4$, with $\omega_A = 1$, $\beta_A = 1/2$, $\beta_B = 2/3$. Dashed red line: standard swap $U_1$ for $\omega_B \in [\beta_A\omega_A/\beta_B,\omega_A] = [3/4,1]$. Purple solid lines: idle swaps $U_2$ for $\omega_B \in [\beta_A\omega_A/2\beta_B,\omega_A/2] = [3/8,1/2]$ and $\tilde{U}_2$ for $\omega_B \in [2\beta_A\omega_A/\beta_B, 2\omega_A] = [3/2,2]$.}
    \label{es6}
\end{figure}

\begin{figure}[H]
    \includegraphics[width=9.5 cm]{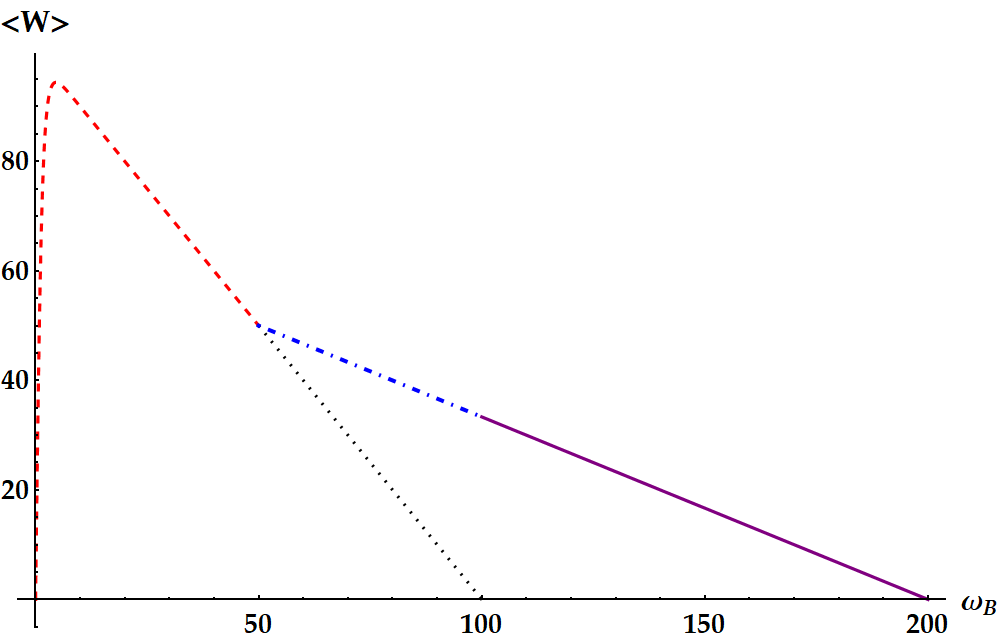}
    \caption{Ergotropy $\langle W \rangle$ as a function of $\omega_B$ in the limiting case $\beta_A/\beta_B=0$, with $\omega_A = 100$, $\beta_A = 0$, $\beta_B = 10$. Blue dot-dashed line: double swap $U_3$. Red dashed line: standard swap $U_1$ inside the corresponding ergotropic regime. Purple solid line: idle swap $\tilde{U}_2$. Black dotted line: standard swap for any $\omega_B$ such that the extracted work is positive.}
    \label{es4}
\end{figure}

Now, we are ready to provide the mean work of Equation~(\ref{maxW}) extracted by each ergotropic transformation. 
In the case of the swap $U_1$, the ergotropy can be expressed in terms of $\omega_A - \omega_B$ units and reads
\vspace{6pt}
\begin{linenomath}
    \begin{equation} \label{w1}
    \begin{aligned}
        \langle W_1\rangle &= 2(\omega_A - \omega_B)\left[\frac{\sinh{\beta_B\omega_B}}{1+2\cosh{\beta_B\omega_B}}-\frac{\sinh{\beta_A\omega_A}}{1+2\cosh{\beta_A\omega_A}}\right] \\
        &= 2(\omega_A - \omega_B)\frac{2\sinh{(\beta_B\omega_B - \beta_A\omega_A)}+\sinh{\beta_B\omega_B}-\sinh{\beta_A\omega_A}}{(1+2\cosh{\beta_A\omega_A})(1+2\cosh{\beta_B\omega_B})}.
        \end{aligned}
    \end{equation}
\end{linenomath}
In the case of the idle swaps $U_2$ and $\tilde{U}_2$, we obtain
\begin{linenomath}
    \begin{equation} \label{w2}
    \begin{aligned}
        \langle W_2 \rangle &= 2(\omega_A - 2\omega_B)\frac{\sinh{\beta_B\omega_B} + \sinh{(\beta_B\omega_B-\beta_A\omega_A)}}{(1+2\cosh{\beta_A\omega_A})(1+2\cosh{\beta_B\omega_B})} 
    \end{aligned}
    \end{equation}
\end{linenomath}
and
\begin{linenomath}
    \begin{equation} \label{w2s}
    \begin{aligned}
        \langle \tilde{W}_2 \rangle &= 2(\omega_B - 2\omega_A)\frac{\sinh{\beta_A\omega_A} + \sinh{(\beta_A\omega_A-\beta_B\omega_B)}}{(1+2\cosh{\beta_A\omega_A})(1+2\cosh{\beta_B\omega_B})}. 
    \end{aligned}
    \end{equation}
\end{linenomath}
Here, we recognize the action described above: the lower frequency qutrit is taken as a qubit whose gap is $2\omega_B$ for $\langle W_2\rangle$ and $2\omega_A$ for $\langle \tilde{W}_2 \rangle$, so that the extracted work is proportional to $\omega_A - 2\omega_B$ and $\omega_B - 2\omega_A$, respectively. As expected, the work extracted from $U_2$ is obtained from the one extracted by $\tilde{U}_2$ just by swapping $A$ with $B$. From Equations~(\ref{w1})--(\ref{w2s}) one also verifies~that
\begin{linenomath}
    \begin{equation}
        \frac{\langle W_1 \rangle}{1-x} = \frac{\langle W_2 \rangle}{1-2x} + \frac{\langle \tilde{W}_2 \rangle}{2-x},
    \end{equation}
\end{linenomath}
where the ratio $x\equiv \omega_B/\omega_A$ is a relevant parameter, as we will find in the following.
\\
In the case of the double swap, we have
\begin{linenomath}
    \begin{equation}  \label{superergo}
    \begin{aligned}
        \langle W_3 \rangle &= 2\frac{\omega_A[\sinh{\beta_B\omega_B} + \sinh{(\beta_B\omega_B - \beta_A\omega_A})]-\omega_B(\sinh{\beta_A\omega_A}+\sinh{\beta_B\omega_B})}{(1+2\cosh{\beta_A\omega_A})(1+2\cosh{\beta_B\omega_B})} \\
        &= \langle W_1\rangle + \frac{1 - 2x}{x - 2} \langle \tilde{W}_2 \rangle = \langle W_2 \rangle + \frac{x}{1-x}\langle W_1\rangle.
    \end{aligned}
    \end{equation}
\end{linenomath}
Here, we see the effects of the atypical behavior of $U_3$: the extracted work is not proportional to any frequency gap. On the contrary, the frequencies $\omega_A$ and $\omega_B$ appear multiplied with different weights. Notice that for $x = 1/2$, one has $\langle W_2\rangle = 0$ and from the second line of Equation~(\ref{superergo}) the double swap $U_3$ extracts the same work as $U_1$, i.e., $\langle W_3 \rangle = \langle W_1 \rangle$. Instead, for $x=1$, namely $\omega_A=\omega_B$, one has $\langle W_1 \rangle = 0$ and $\langle W_3 \rangle = \langle \tilde{W}_2 \rangle$. Finally, for $x=2$, we have $\langle \tilde{W}_2 \rangle = 0$ and again $\langle W_3 \rangle = \langle W_1 \rangle$. 
In Figure~\ref{wereg1}, we represent the ergotropy extraction in the case $\beta_A/\beta_B \in (0,1/4)$. In particular, we set the ratio $\beta_A/\beta_B = 1/16$, as in Figure~\ref{case1}, with $\beta_B = 10$. Note that the pretended discontinuities in the transitions between different ergotropic regions are just cusps, as it can be recognized in Figures~\ref{es2}--\ref{es4}.

In these figures, we show specific examples of ergotropy extraction as a function of $\omega_B$, by fixing all the other parameters. Figure~\ref{es2} displays the case $\beta_A/\beta_B < 1/4$, with $\beta_A/\beta_B = 1/8$. Therefore, this is not a critical point, and for varying $\omega_B$, we span all the non-equivalent ergotropic transformations. The black dot line displays the work extracted from the standard swap $U_1$ so that we can see how it is outperformed by the other unitaries outside its own ergotropic regime. Moreover, the solid lines, corresponding to $U_2$ and $\tilde{U}_2$, show that the regime of operation of an ergotropic heat engine is enlarged with respect to the swap Otto engine. In Figure~\ref{es3}, we consider the critical point $\beta_A/\beta_B = 1/4$, which represents the transition between the cases in Figures~\ref{case1} and~\ref{case2}, where the ergotropic regime of the standard swap vanishes. Indeed, here we do not have any ergotropic contribution from $U_1$, except for the limiting case $\omega_A = 2\omega_B$, where the work extracted coincides with the one provided by $U_3$, identified by the red point. In Figure~\ref{es1}, we show the ergotropy as a function of $\omega_B$ for the critical point $\beta_A/\beta_B = 1/2$, which is the transition point between the cases of Figures~\ref{case2} and~\ref{case3}. As expected, the double swap $U_3$ is never required to extract the ergotropy. The maximum work is extracted by the idle swap $U_2$ for $\omega_B < \omega_A/2$, by the standard swap $U_1$ for $\omega_A/2 < \omega_B < \omega_A$ and by $\tilde{U}_2$ for $\omega_A < \omega_B < 2\omega_A$. For the case $\beta_A/\beta_B > 1/2$ of Figure~\ref{case3}, we fix in Figure~\ref{es6} $\beta_A/\beta_B = 3/4$. As in the previous case, $U_3$ is not needed and, furthermore, there are two more passive regions. Finally, in the last example in Figure~\ref{es4}, we plot the ergotropy for the ideal case $\beta_A/\beta_B = 0$, by setting $\beta_A$ to $0$ and finite large values for $\omega_A$ and $\beta_B$. In particular, the high value of $\omega_A$ allows to see that the extracted work is large when $\omega_A - \omega_B$ is large, except for the limiting case $\omega_B \rightarrow 0$ (in such a case indeed we would have $\beta_A\omega_A = \beta_B\omega_B = 0$, implying $\langle W_1 \rangle =0$). 

\unskip
\unskip
\unskip
\unskip
In summary, in the regime of operation of the standard swap Otto engine, i.e., $\omega_A > \omega_B \wedge \beta_A\omega_A < \beta_B\omega_B$, the work extraction may be improved by replacing the swap $U_1$ with the permutation $U_3$. Moreover, the idle swaps $U_2$ and $\tilde{U}_2$ even allow to enlarge the range of operation of the heat engine.

\subsection{Entropy Production} \label{r3}
Let us now evaluate the mean entropy production of the quantum heat engine in order to study its relation with the work fluctuations and to explore the validity or violation of TURs. As mentioned in Section~\ref{mm}, the mean entropy production is given by
\begin{linenomath}
    \begin{equation}
    \langle \Sigma \rangle = (\beta_A - \beta_B)\langle \Delta E_A \rangle - \beta_B \langle W \rangle.
\end{equation}
\end{linenomath}
We can evaluate the moments of $W$ and $\Delta E_A$ through the derivatives of the characteristic function, according to Equations~(\ref{tf}) and (\ref{mom}) of Appendix~\ref{appA1}. Due to the conservation laws for $U_1$, $U_2$, and $\tilde{U}_2$ as in Equations~(\ref{comm1}) and~(\ref{comm2}), according to Equation~(\ref{sym}), we have
\begin{linenomath}
    \begin{equation} \label{tsprop}
        \langle W^l\Delta E_A^s\rangle = \alpha^s \langle W^{l+s} \rangle,
    \end{equation}
\end{linenomath}
where $\alpha = \omega_A/(\omega_B - \omega_A)$ for $U_1$, $\alpha = 2\omega_B/(\omega_A - 2\omega_B)$ for $U_2$, and $\alpha = 2\omega_A/(\omega_B - 2\omega_A)$ for $\tilde{U}_2$. Hence, the entropy production of $U_1$, $U_2$, and $\tilde{U}_2$ is proportional to their pertaining work, and one has
\begin{linenomath}
    \begin{equation} \label{s1}
    \begin{aligned}
        \langle \Sigma_1 \rangle 
         &= \frac{\beta_B\omega_B -\beta_A\omega_A}{\omega_A - \omega_B}\langle W_1 \rangle, \\
    \langle \Sigma_2 \rangle &= \frac{2\beta_B\omega_B-\beta_A\omega_A}{\omega_A - 2\omega_B}\langle W_2\rangle, \\
    \langle \tilde{\Sigma}_2 \rangle &= \frac{\beta_B\omega_B - 2\beta_A\omega_A}{2\omega_A-\omega_B}\langle \tilde{W}_2\rangle,
    \end{aligned}
\end{equation}
\end{linenomath}
where $\langle W_1\rangle$, $\langle W_2\rangle$ and $\langle \tilde{W}_2\rangle$ are given in Equations~(\ref{w1}),~(\ref{w2}) and~(\ref{w2s}), respectively.

Equation~(\ref{tsprop}) does not hold for $U_3$, and the entropy production explicitly is given by
\begin{equation} \label{s3}
\begin{aligned}
    \langle \Sigma_3 \rangle &= 2 \frac{\beta_B\omega_B(\sinh{\beta_A\omega_A} +\sinh{\beta_B\omega_B})-\beta_A\omega_A[\sinh{\beta_B\omega_B}+\sinh{(\beta_B\omega_B-\beta_A\omega_A})]}{(1+2\cosh{\beta_A\omega_A})(1+2\cosh{\beta_B\omega_B})}.
\end{aligned}
\end{equation}
Note that in all cases the mean entropy production is positive and depends only on the ratios between frequency and temperature and not on the bare frequencies.

\subsection{Work Distribution} \label{r4}

\begin{figure}[H]
    \includegraphics[width=8.5 cm]{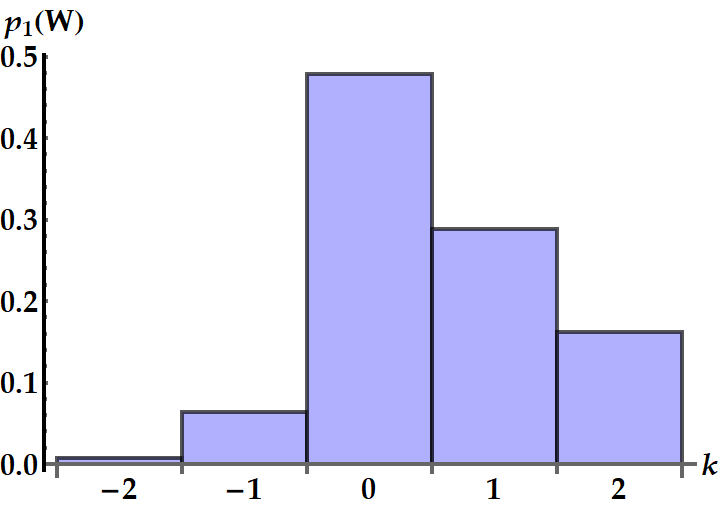}
    \caption{Distribution $p_1(W=k(\omega_A-\omega_B))$ of the work extracted by the standard swap $U_1$ in $\omega_A - \omega_B$ units. We set $\beta_A\omega_A = 0.5$ and $\beta_B\omega_B = 2$.}
    \label{p1w}
\end{figure}

We can now provide the explicit expression for the distribution of work $p(W)$ pertaining to each ergotropic transformation. As shown in Appendix~\ref{appA1} (see 
 Equation~(\ref{pw})), we~have
\begin{linenomath} 
    \begin{equation} \label{gendist}
    p(W) = \sum_{n,m,l,s} p_{n,m} q(l,s|n,m) \delta(W-\omega_A(n-l)-\omega_B(m-s))
\end{equation}
\end{linenomath}
where $p_{n,m}$ is the energy distribution of the input state, namely
\begin{linenomath}
    \begin{equation}
    p_{n,m} = \frac{1}{Z_AZ_B}e^{-\beta_A\omega_An}e^{-\beta_B\omega_Bm}
\end{equation}
\end{linenomath}
while $q(l,s|n,m)$ is the energy conditional distribution after the evolution $U$, given the input energy levels $n$ and $m$, i.e.,
\begin{linenomath}
    \begin{equation}
    q(l,s|n,m) = |\langle l,s|U|n,m\rangle|^2.
\end{equation}
\end{linenomath}
In the case of the standard swap $U_1$, the conditional distribution reads $q_1(l,s|n,m) = \delta_{l,m}\delta_{n,s}$, and hence
\begin{linenomath}
    \begin{equation}
\begin{aligned}
    p_1(W) &= \sum_{n,m=0}^{2} p_{n,m} \delta(W-(n-m)\omega_A-(m-n)\omega_B), 
\end{aligned}
\end{equation}
\end{linenomath}
which is a 5-point distribution. Explicitly, upon naming $k\equiv n-m$, one has
\begin{linenomath}
    \begin{equation} \label{pu1}
\begin{aligned}
    &p_1(W=k(\omega_A-\omega_B)) = \\
    =& \frac{1}{Z_AZ_B}\frac{1-\exp{[-(k+3)(\beta_A\omega_A+\beta_B\omega_B)]}}{1-\exp{[-(\beta_A\omega_A+\beta_B\omega_B)]}}e^{\beta_A\omega_A k} \quad \quad \text{with} \quad k\in[-2,0) \\
    =& \frac{1}{Z_AZ_B}\frac{1-\exp{[(k-3)(\beta_A\omega_A+\beta_B\omega_B)]}}{1-\exp{[-(\beta_A\omega_A+\beta_B\omega_B)]}}e^{-\beta_B\omega_B k} \quad \quad \text{with} \quad k\in[0,2].
\end{aligned}
\end{equation}
\end{linenomath}
A specific example is plotted in Figure~\ref{p1w}. Equation~(\ref{pu1}) is consistent with the general result given in Ref.~\cite{max} for the work distribution in swap engines based on two qudits.

\begin{figure}[H]
    \includegraphics[width=8.5 cm]{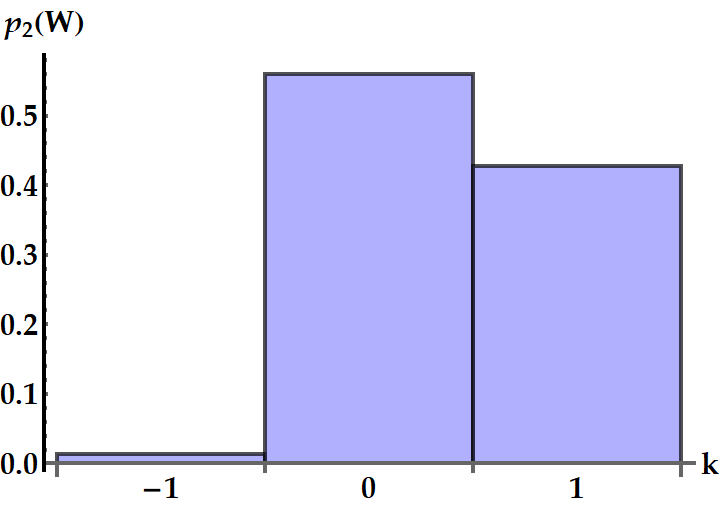}
    \caption{Distribution $p_2(W=k(2\omega_B-\omega_A))$ of the work extracted by the idle swap $U_2$ in $2\omega_B - \omega_A$ units. We set $\beta_A\omega_A = 0.5$ and $\beta_B\omega_B = 2$.}
    \label{p2w}
\end{figure}

\begin{figure}[H]
    \includegraphics[width=8.5 cm]{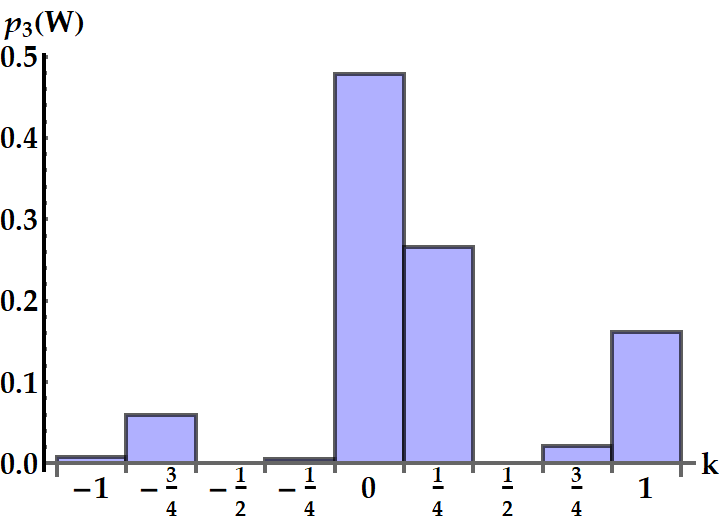}
    \caption{Distribution $p_3(W=k\omega_A))$ of the work extracted by the double swap $U_3$ in $\omega_A$ units in the case $\omega_B/\omega_A = 3/4$. We set $\beta_A\omega_A = 0.5$ and $\beta_B\omega_B = 2$.}
    \label{p3w}
\end{figure}

Now, we focus on the idle swap $U_2$. Due to its asymmetric action on systems $A$ and $B$, the conditional distribution is slightly more complicated and reads
\begin{linenomath}
    \begin{equation}
    \begin{aligned}
        q_2(l,s|n,m) = \sum_{m=0}^2\delta_{n,m}\delta_{l,s}\delta_{s,m} + \delta_{s,2-m}(\delta_{n, m\bigoplus 1} + \delta_{n, m\bigoplus 2}) +(\delta_{l, s\bigoplus 1} + \delta_{l, s\bigoplus 2}) 
    \end{aligned}
\end{equation}
\end{linenomath}
where $\bigoplus$ denotes the sum $\text{mod}\, 3$. Hence, one retrieves the following 3-point distribution
\begin{linenomath}
  \begin{equation} \label{p2}
    \begin{aligned}
    p_2(W) =& \left(\sum_{n=0}^2 p_{n,n} + p_{01} + p_{21}\right) \delta(W) + (p_{10}+p_{20})\delta(W-\omega_A + 2\omega_B) \\
    & + (p_{02}+p_{12})\delta(W+ \omega_A - 2\omega_B).
\end{aligned}
\end{equation}
\end{linenomath}
An example is depicted in Figure~\ref{p2w}.
Similarly, in the case of $\tilde{U}_2$ one has
\begin{linenomath}
    \begin{equation} \label{p2sym}
\begin{aligned}
    \tilde{p}_2(W) =& \left(\sum_{n=0}^2 p_{n,n} + p_{10} + p_{12}\right) \delta(W) + (p_{01}+p_{02})\delta(W-\omega_B + 2\omega_A) \\
    & + (p_{20}+p_{21})\delta(W+\omega_B - 2\omega_A).
\end{aligned}
\end{equation}
\end{linenomath}
Finally, since $U_3 = U_2 U_1$, for the double swap we readily find
\begin{linenomath}
    \begin{equation}
    q_3(l,s|n,m) = q_2(l,s|m,n),
\end{equation}
\end{linenomath}
and then one obtains the following 7-point distribution
\begin{linenomath}
    \begin{equation}
    \begin{aligned}
        p_3(W) =& \sum_{n=0}^2 p_{n,n}\delta(W) + p_{10}\delta(W - \omega_A + \omega_B)+ p_{12}\delta(W + \omega_A - \omega_B) + \\
        &+  p_{01}\delta(W+\omega_B) + p_{21}\delta(W-\omega_B) + p_{02}\delta(W+\omega_A) + p_{20}\delta(W-\omega_A).
    \end{aligned}
\end{equation}
\end{linenomath}
A specific example is provided in Figure~\ref{p3w}.

\subsection{Work Fluctuations and TURs} \label{r5}
Here, we evaluate the relative fluctuations of the work extracted by the ergotropic transformations and compare them to the lower bounds identified by different thermodynamic uncertainty relations (TURs).

We 
 can find the relative fluctuations as the ratio between the variance of the extracted work and the square of its mean value, namely $\text{var}(W)/\langle W\rangle^2 = \langle W^2 \rangle/\langle W\rangle^2 - 1$, with $\text{var}(W) = \langle W^2 \rangle - \langle W \rangle^2$ . The second moment of the extracted work can be obtained from the characteristic function as in Equations~(\ref{mom}) and~(\ref{a5}), and one has
\begin{linenomath}
\begin{equation}
\begin{aligned}
    \langle W^2_k \rangle =& \text{Tr}[(H_A\otimes \mathbb{I}_B + \mathbb{I}_A\otimes H_B)^2\rho_0] + \text{Tr}[(H_A\otimes \mathbb{I}_B + \mathbb{I}_A\otimes H_B)^2U_k\rho_0U_k^{\dagger}] \\
    &- 2\text{Tr}[U_k^{\dagger}(H_A\otimes \mathbb{I}_B + \mathbb{I}_A\otimes H_B)U_k(H_A\otimes \mathbb{I}_B + \mathbb{I}_A\otimes H_B)\rho_0].
\end{aligned}
\end{equation}
\end{linenomath}
For the standard swap $U_1$ one obtains
\begin{linenomath}
    \begin{equation} \label{flut1}
    \begin{aligned}
    \frac{\text{var}(W_1)}{\langle W_1 \rangle^2} =& (1+2\cosh{\beta_A\omega_A})(1+2\cosh{\beta_B\omega_B}) \\ 
    &\times\frac{\cosh{\beta_A\omega_A}+\cosh{\beta_B\omega_B}+4\cosh{(\beta_B\omega_B-\beta_A\omega_A)}}{2[\sinh{\beta_B\omega_B}-\sinh{\beta_A\omega_A}+2\sinh{(\beta_B\omega_B-\beta_A\omega_A)}]^2} - 1
    \end{aligned}
\end{equation}
\end{linenomath}
which is in agreement with the general result of the swap engine with two qudits of Ref.~\cite{max}. As expected, the fluctuations of the standard swap are invariant under swapping $A$ and $B$.
\\
For the idle swap $U_2$, we have
\begin{linenomath}
    \begin{equation} \label{fl2}
    \scalebox{1.05}[1.05]{$\frac{\text{var}(W_2)}{\langle W_2 \rangle^2} = (1+2\cosh{\beta_A\omega_A})(1+2\cosh{\beta_B\omega_B}) \frac{\cosh{\beta_B\omega_B}+\cosh{(\beta_B\omega_B-\beta_A\omega_A)}}{2[\sinh{\beta_B\omega_B}+\sinh{(\beta_B\omega_B-\beta_A\omega_A)}]^2} - 1.$}
\end{equation}
\end{linenomath}
As for the ergotropy in Equations~(\ref{w2}) and~(\ref{w2s}), the expression for $\text{var}(\tilde{W}_2)/\langle\tilde{W}_2 \rangle^2$ is simply obtained by exchanging $A$ with $B$ in Equation~(\ref{fl2}).
Note that the fluctuations of both the standard and the idle swap depend only on the products $\beta\omega$.

This 
 is not the case for the double swap $U_3$, which depends also on the frequency ratio $x=\omega_B/\omega_A$ as follows
\begin{linenomath}
    \begin{equation} \label{fluct3}
    \begin{aligned}
    \frac{\text{var}(W_3)}{\langle W_3 \rangle^2} =& (1+2\cosh{\beta_A\omega_A})(1+2\cosh{\beta_B\omega_B}) \\
    &\times\frac{x^2\cosh{\beta_A\omega_A}+(1-x)^2\cosh{\beta_B\omega_B}+\cosh{(\beta_B\omega_B-\beta_A\omega_A)}}{2\left[(1-x)\sinh{\beta_B\omega_B}-x\sinh{\beta_A\omega_A}+\sinh{(\beta_B\omega_B-\beta_A\omega_A)}\right]^2} - 1.
    \end{aligned}
\end{equation}
\end{linenomath}
For all the ergotropic transformations the fluctuations are minimized in the limiting case where $\beta\omega\rightarrow 0$ for one qutrit and $\beta\omega\rightarrow\infty$ for the other one. In the case of the swap, being naturally invariant under swap symmetry, we can either set $\beta_A\omega_A$ to zero and $\beta_B\omega_B$ to infinity or the other way around. On the contrary, the case of the idle and the double swap is asymmetric and we achieve the minimum of the fluctuations for $\beta_B\omega_B \rightarrow 0 \wedge \beta_A\omega_A \rightarrow \infty$ in the case of $\tilde{U}_2$ and for $\beta_A\omega_A \rightarrow 0 \wedge \beta_B\omega_B \rightarrow \infty$ in the case of $U_2$ and $U_3$. Here, we mainly focus on the transformations that extract the ergotropy in the same poset identified by the products $\beta\omega$. In particular, we choose the poset defined by $\beta_A\omega_A < \beta_B\omega_B$, where the optimal evolutions are $U_1$, $U_2$ and $U_3$.
In the case of the double swap $U_3$, the minimization has to be
performed also on the frequency ratio and the infimum is obtained for
$x \rightarrow 0$.
The optimization of the fluctuations over the whole span of the parameters readily provides
\begin{linenomath}
    \begin{equation} \label{boundcasc}
       \frac{2}{3}  = \inf_{\beta_A\omega_A, \beta_B\omega_B}\frac{\text{var}(W_1)}{\langle W_1 \rangle^2} >  \inf_{\beta_A\omega_A, \beta_B\omega_B}\frac{\text{var}(W_2)}{\langle W_2 \rangle^2} = \inf_{\beta_A\omega_A, \beta_B\omega_B, x}\frac{\text{var}(W_3)}{\langle W_3 \rangle^2} = \frac{1}{2}.
    \end{equation}
\end{linenomath}
Therefore, it turns out that $U_2$ and $U_3$ achieve smaller fluctuations than $U_1$. 

We now investigate if damping the noise comes together with the
extraction of the ergotropy. While for $U_1$ this is always the case,
the same is not true for $U_2$ and $U_3$. The idle swap $U_2$ extracts
the ergotropy for $\beta_B\omega_B < \beta_A\omega_A$, where the
condition for the minimization of fluctuations corresponding to $U_2$
does not hold. Interestingly, in that region, it is $\tilde{U}_2$ the
ergotropy extractor. Within the ergotropic region of $U_2$, we need to
take $\beta_B\omega_B \rightarrow 0 \wedge \beta_A\omega_A \rightarrow
\infty$, which provides $\text{var}(W_2)/\langle W_2 \rangle^2 =
2$. For $U_3$, on the contrary, the condition on the ratios
$\beta\omega$ for optimal fluctuations is compatible with the
extraction of ergotropy, but with the additional constraint $x \geq
1/2$. The minimization over $x$ then sets it to $1/2$, and, as
discussed after Equation~(\ref{superergo}), for that frequency ratio
$\langle W_3\rangle = \langle W_1\rangle$.  To sum up, if we aim to
optimize the noise inside the ergotropic regimes of each ergotropic
transformation, we find that the best performance is achieved by the
standard swap since
\begin{linenomath}
    \begin{equation} \label{boundcasc2}
       2 = \inf_{\beta_A\omega_A, \beta_B\omega_B}\frac{\text{var}(W_2)}{\langle W_2 \rangle^2} > \inf_{\beta_A\omega_A, \beta_B\omega_B}\frac{\text{var}(W_1)}{\langle W_1 \rangle^2} =  \inf_{\beta_A\omega_A, \beta_B\omega_B, x}\frac{\text{var}(W_3)}{\langle W_3 \rangle^2} = \frac{2}{3}.
    \end{equation}
\end{linenomath}
\\
In this last regime where ergotropy extraction and minimal noise coexist, we finally note that the standard swap extracts more work than the idle and double swap. In fact, one has
\begin{linenomath}
    \begin{equation}
    \begin{aligned}
    &\sup_{\omega_B} W_1(\beta_A\omega_A\rightarrow 0, \beta_B\omega_B \rightarrow \infty) = \omega_A, \\
    &\sup_{\omega_B} W_2(\beta_A\omega_A \rightarrow \infty, \beta_B\omega_B\rightarrow 0) = \frac{\omega_A}{3}, \\
    &\sup_{\omega_B} W_3(\beta_A\omega_A\rightarrow 0, \beta_B\omega_B \rightarrow \infty) = \frac{\omega_A}{2}.
    \end{aligned}
\end{equation}
\end{linenomath}
We remark that the results found so far do imply that the standard swap is the best operation in terms of fluctuations and extracted work in the optimal limiting case $\beta_A\omega_A\rightarrow 0 \wedge \beta_B\omega_B \rightarrow \infty$, but the same does not hold for intermediate values of $\beta\omega$, as we shall see in the following.

\begin{figure}[H]
    \includegraphics[width=10.5 cm]{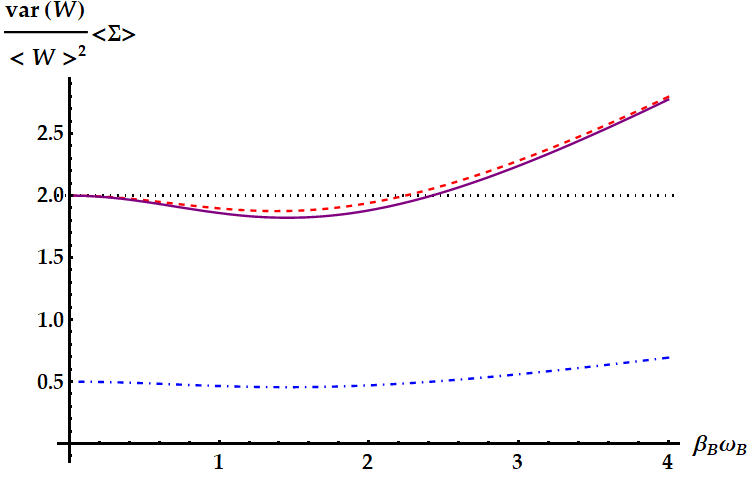}
    \caption{Product 
 of the relative fluctuations with the mean entropy production, which is lower bounded by 2 (black dotted line) in the standard TUR of Equation~(\ref{fluct}) and by a function of the mean entropy in Equation~(\ref{landi}). We set $\beta_A\omega_A = 10^{-3}$. The plot shows the violations due to the standard swap $U_1$ (red dashed line), the idle swap $U_2$ (solid purple line) and the double swap $U_3$ (blue dot-dashed line) as a function of $\beta_B\omega_B$.}
    \label{stur}
\end{figure}

Now, we compare the relative fluctuations of the ergotropic engine in asymptotic and non-asymptotic cases with the bounds derived from the most significant TURs. We recall that the double swap $U_3$ is not Hermitian. Therefore, as remarked in the Appendix after Equation~(\ref{bf}), $U_3$ could violate all the TURs based on the equivalence between forward and backward processes. On the other hand, we already know from previous works~\cite{max} that the swap itself breaks the standard TUR in Equation~(\ref{fluct}). We study in Figure~\ref{stur} the violation of the standard TUR as a function of $\beta\omega$ in conditions of minimal fluctuations, independently from the ergotropic regime. Namely, in the case of $U_1$ (red dashed line), $U_2$ (purple solid line) and $U_3$ (blue dot-dashed line) the free variable is $\beta_B\omega_B$ with $\beta_A\omega_A \ll 1$. Just for $U_3$, we also need $\omega_B/\omega_A \ll 1$. We remark that here we are not focusing on the ergotropy extraction, but only on the properties of the evolutions $U_1$, $U_2$, and $U_3$ in terms of work fluctuations. We find that all three ergotropic transformations break the standard thermodynamic uncertainty relation. In particular, the violation due to $U_3$ is impressive. As found in~\cite{max}, when the evolution is the standard swap the relative fluctuations for the extracted work satisfies
\begin{linenomath}
    \begin{equation} \label{maxtur}
        \frac{\text{var}(W)}{\langle W \rangle^2} \geq \frac{2}{\langle \Sigma \rangle} - 1.
    \end{equation}
\end{linenomath}

The variation of Equation~(\ref{maxtur}) from the standard TUR explains the slight violation found in Figure~\ref{stur}, where the lower bound from the standard TUR is displayed as a black dotted line. Similarly to $U_1$, also $U_2$ and $\tilde{U}_2$ satisfy Equation~(\ref{maxtur}). In fact,
\begin{linenomath}
    \begin{equation}
\begin{aligned}
    \frac{\langle W_2^2 \rangle}{\langle W_2 \rangle^2} = \frac{f(\beta_A\omega_A,\beta_B\omega_B)}{\langle \Sigma_2\rangle} 
\end{aligned}
\end{equation}
\end{linenomath}
where 
\begin{linenomath}
    \begin{equation}
        f(x,y) \equiv (2y - x) \frac{\cosh{y} + \cosh{(y-x)}}{\sinh{y}+\sinh{(y-x)}},
    \end{equation}
\end{linenomath}
which satisfies
\begin{linenomath}
    \begin{equation}
    f(x,y) \geq 2 \quad \forall\, x, y\geq 0.
\end{equation}
\end{linenomath}
The fluctuations originated from $U_3$, instead, can break the TUR in Equation~(\ref{maxtur}). Such violation stems from the asymmetry of the process described by $U_3$, which is not Hermitian. Indeed, we note that a necessary condition for the TURs in Equations~(\ref{fluct}) and~(\ref{maxtur}) to hold is the equivalence between forward and backward process, i.e., $p_{B}(W,\Delta E_A)=p(W,\Delta E_A)$. Moreover, note that the double swap is the only transformation whose fluctuations depend also on the frequency ratio while leaving the mean entropy as a function of just $\beta_A\omega_A$ and $\beta_B\omega_B$. Therefore, in this case, we can optimize over a third parameter without changing the lower bound of the TUR.

\begin{figure}[H]
    \includegraphics[width=10.5 cm]{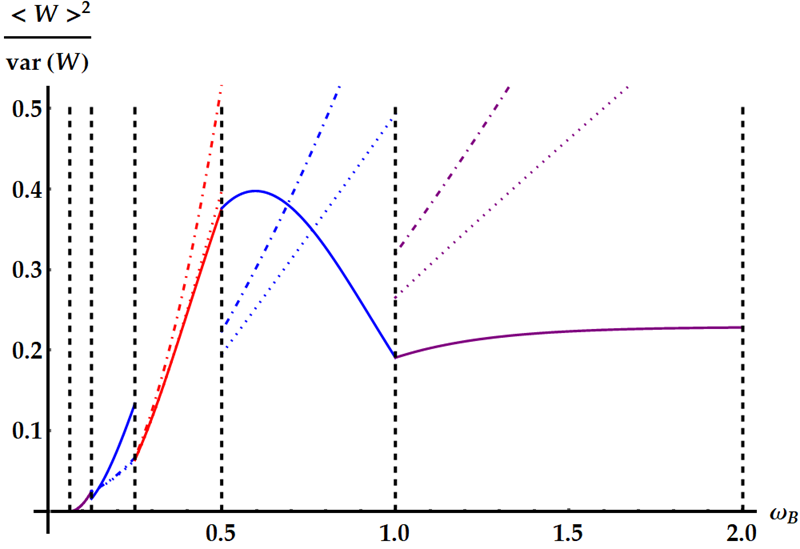}
    \caption{SNRs obtained from the ergotropy extraction of the example and parameters as in Figure~(\ref{es2}). The vertical dashed lines separate different the ergotropic regimes. We have the idle swaps $U_2$ for $\omega_B \in [1/16,1/8]$ and $\tilde{U}_2$ for $\omega_B\in[1,2]$ (purple lines), the double swap $U_3$ for $\omega_B \in [1/8,1/4]$ and for $\omega_B \in [1/2,1]$ (blue lines) and $U_1$ for $\omega_B \in [1/4,1/2]$ (red line). The solid curves display the SNR. The dotted and dot-dashed lines show the upper bounds provided by the standard TUR in Equation~(\ref{fluct}) and the tight TUR in Equation~(\ref{landi}). In the region $\omega_B \in [1/2,1]$, corresponding to the non-Hermitian unitary $U_3$, strong violations of both TURs are apparent.}
    \label{flutt_es2}
\end{figure}

The violation of the TUR in Equation~(\ref{maxtur}) by $U_3$ can also be found in realistic cases, i.e., even if we do not set the parameters to the values minimizing the fluctuations. Actually, these cases are the most relevant to be considered, not only because closer to experimental applications but especially because they keep into account the ergotropy extraction provided by the different evolutions. For instance, consider the case of Figure~\ref{es2}, where $\omega_A = 1$, $\beta_A = 0.5$, $\beta_B = 4$ and $\omega_B$ is left free. Correspondingly, in Figure~\ref{flutt_es2}, we plot the signal-to-noise ratio (SNR) of the extracted work for each transformation in its ergotropic regime, together with the lower bound of Equation~(\ref{fluct}) (dotted lines). Firstly, note that the double swap violates the TUR even if we are far from the optimal conditions on the parameters maximizing the SNR. Second, the TUR is violated in both regimes where $U_3$ extracts the ergotropy ($\omega_B \in [1/8,1/4] \cup [1/2,1]$). Third, differently from what we found in the case of optimal conditions, $U_2$ and $U_3$ can achieve better SNRs than the standard swap $U_1$ where the ergotropy is extracted.

The standard TUR is not the only relevant lower bound which we show in Figure~\ref{flutt_es2}. The tightest TUR that cannot be violated by any time-symmetric process was found in Ref.~\cite{landi} and, applied to the extracted work, reads
\begin{linenomath}
    \begin{equation} \label{landi}
        \frac{\text{var}(W)}{\langle W \rangle^2} \geq \text{csch}^2[g(\langle\Sigma\rangle/2)]
    \end{equation}
\end{linenomath}
where $g(x)$ is the inverse of the function $x\tanh(x)$. Therefore, we expect that neither $U_1$ nor $U_2$ can violate this TUR, while $U_3$ in principle can. This is what we see in Figure~\ref{flutt_es2}, where the dot-dashed lines correspond to the lower bound determined by Equation~(\ref{landi}): the SNR identified by the double swap $U_3$ is the only one that can violate the tight TUR, also within its ergotropic regime.

\begin{figure}[H]
    \includegraphics[width=11 cm]{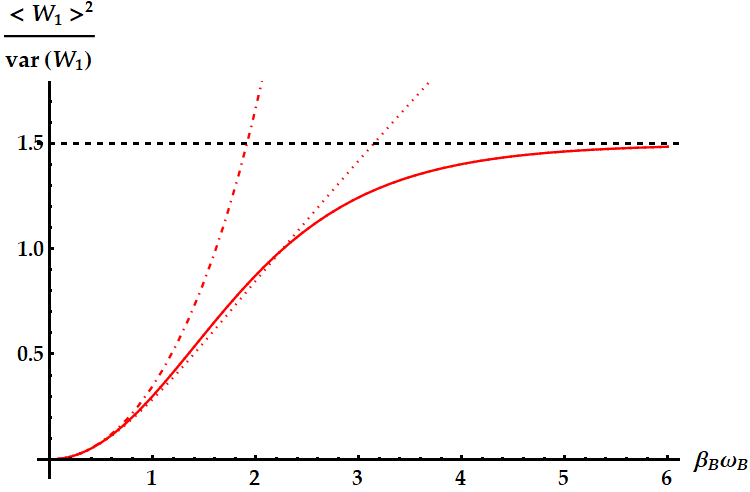}
    \caption{SNR of the work extracted by the standard swap $U_1$ (solid line) in ideal optimal conditions, with $\beta_A\omega_A = 10^{-3}$. The dotted and dot-dashed lines display the upper bound from the standard TUR in Equation~(\ref{fluct}) and the tight TUR in Equation~(\ref{landi}), respectively. The dashed horizontal line highlights the asymptotic limit of the SNR.}
    \label{snr1}
\end{figure}

\begin{figure}[H]
    \includegraphics[width=10.5 cm]{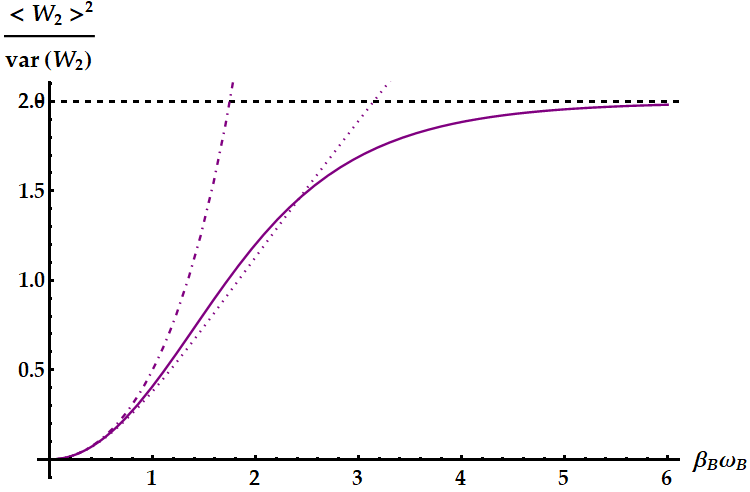}
    \includegraphics[width=10.5 cm]{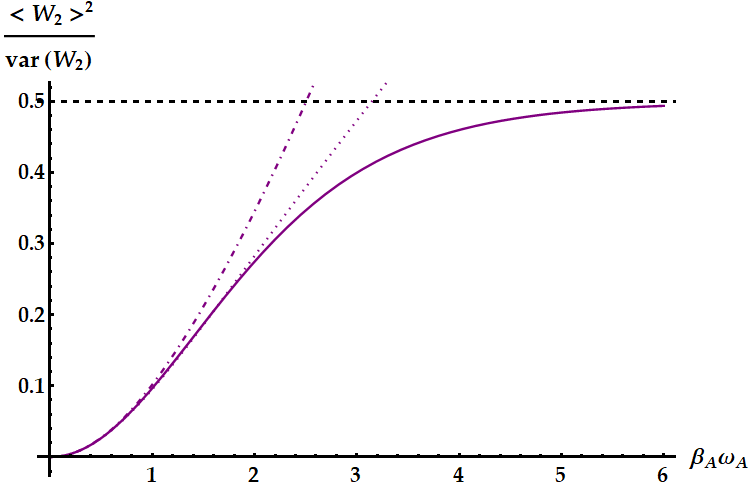}
    \caption{SNR of the work extracted by the idle swap $U_2$ (solid lines). The dotted lines display the upper bound from the standard TUR in Equation~(\ref{fluct}), while the dot-dashed lines display the upper bound from the tight TUR in Equation~(\ref{landi}). The dashed horizontal lines highlight the limit of the SNR. (\textbf{Up panel)}: conditions for the maximum SNR independently from the ergotropy extraction, namely $\beta_B\omega_B > \beta_A\omega_A\sim 0$. Here, we set $\beta_A\omega_A = 10^{-3}$. (\textbf{Bottom panel}): conditions for the maximum SNR within the ergotropic regime of $U_2$, namely $\beta_A\omega_A > \beta_B\omega_B\sim 0$. Here we set $\beta_B\omega_B = 10^{-3}$.}
    \label{snr2}
\end{figure}

\begin{figure}[H]
    \includegraphics[width=10.5 cm]{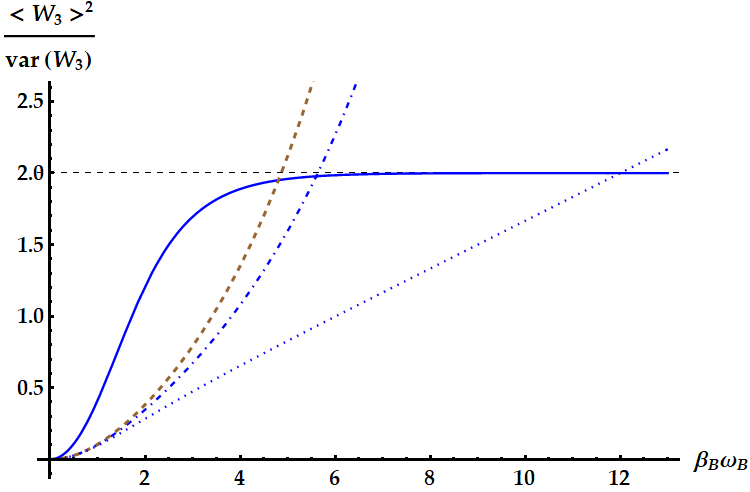}
    \includegraphics[width=10.5 cm]{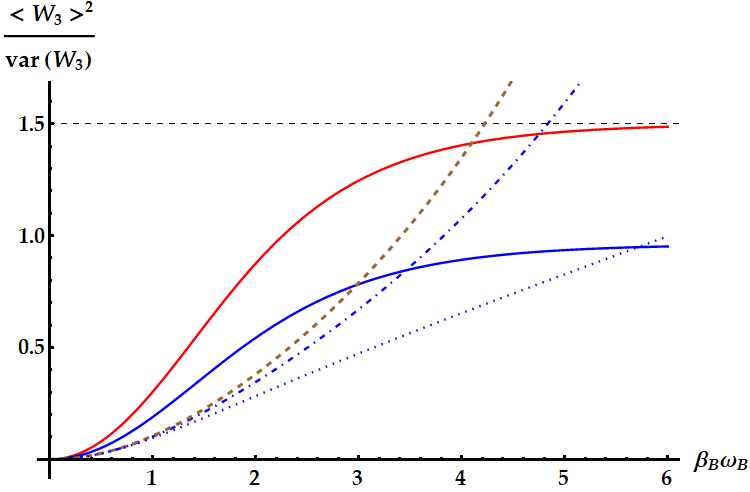}
    \caption{SNR of the work extracted by the double swap $U_3$. The dotted and dot-dashed lines display the upper bound from the standard TUR in Equation~(\ref{fluct}) and the tight TUR in Equation~(\ref{landi}). The dashed brown lines display the bound from the loosest TUR for time-symmetric processes in Equation~(\ref{loose}). The dashed horizontal lines highlight the limit of the SNR. (\textbf{Up panel}): conditions for the maximum SNR independently from the ergotropy extraction, namely $\beta_B\omega_B > \beta_A\omega_A \sim 0$ and $\omega_B/\omega_A \sim 0$. Here, we set $\beta_A\omega_A = \omega_B/\omega_A = 10^{-3}$. The solid blue line displays the SNR. (\textbf{Bottom panel}): conditions for the maximum SNR within the ergotropic regime of $U_3$, namely $\beta_B\omega_B > \beta_A\omega_A\sim 0$ and $\omega_B/\omega_A \in [1/2,1)$. Here, we set $\beta_A\omega_A = 10^{-3}$ and show the cases obtained from two different choices of the frequency ratio. The red line displays the choice optimizing the SNR, i.e., $\omega_B/\omega_A = 1/2$, which reduces the statistics of the work extracted by the double swap to the one extracted by the standard swap. The blue solid line displays the case $\omega_B/\omega_A = 3/4$.}
    \label{snr3}
\end{figure}

We focus more in detail on the violation of the TURs above in Figures~\ref{snr1}--\ref{snr3}, where we plot the SNRs for the three evolutions both for optimal values of the parameters independently from the ergotropy extraction and within the corresponding ergotropic regime.
In particular, Figure~\ref{snr1} displays the performance of the standard swap $U_1$. Here, we set $\beta_A\omega_A\ll 1$, which implies that the fluctuations are minimized for large $\beta_B\omega_B$. As $\beta_B\omega_B$ increases, the signal-to-noise ratio approaches the inverse of the minimal fluctuations, namely $3/2$, in agreement with Equation~(\ref{boundcasc}). Again, we find a slight violation of the standard TUR (dotted line).

In Figure~\ref{snr2}, we show the performance of the idle swap $U_2$ where it maximizes the SNR (first panel) and extracts the ergotropy (second panel). Therefore, in the former case, we set $\beta_A\omega_A\ll 1$ and retrieve the optimization of the SNR for large values of $\beta_B\omega_B$, as in Equation~(\ref{boundcasc}). In the regime where $U_2$ extracts the ergotropy, as in Equation~(\ref{boundcasc2}), we find an optimal SNR approaching $1/2$ for $\beta_A\omega_A \gg \beta_B\omega_B\sim 0$ and an almost negligible violation of the standard TUR.
Neither the standard nor the idle swap violates the tight TUR in Equation~(\ref{landi}), displayed as a dashed-dotted line.

The case of the double swap, displayed in Figure~\ref{snr3}, is radically different. If we neglect the conditions for the ergotropy extraction, here we can optimize also over the frequency ratio $\omega_B/\omega_A$ and we can set it to zero, while $\beta_A\omega_A\sim 0$, implying that we expect to find the optimal SNR for large $\beta_B\omega_B$, as in Equation~(\ref{boundcasc}). Again, the standard TUR is violated, but, compared with the previous cases, the corresponding bound is saturated for larger values of $\beta_B\omega_B$, where the SNR approaches its maximum. Most importantly, the tight TUR of Equation~(\ref{landi}) is also violated, both when the SNR is optimized (first panel) and when the ergotropy is extracted (second panel).

We also compare the SNR of $U_3$ with the loosest bound that always holds for time-symmetric processes~\cite{hase,pros,mer} given by
\begin{linenomath}
    \begin{equation} \label{loose}
        \frac{\text{var}(W)}{\langle W \rangle^2} \geq \frac{2}{e^{\langle\Sigma\rangle} - 1}.
    \end{equation}
\end{linenomath}
The bound from Equation~(\ref{loose}) is displayed as a brown line in Figure~\ref{snr3}. The violation that we find is a consequence of the fact that $U_3$ is not Hermitian.

In the second panel of Figure~\ref{snr3}, as mentioned above, we explore the performance of the double swap $U_3$ in its ergotropic regime, where $\omega_B/\omega_A \in [1/2,1]$. The best performance is obtained for $\omega_B/\omega_A = 1/2$, where the amount of work extracted by $U_3$ is the same as the one extracted by $U_1$ (red line in Figure~\ref{snr3}). We also plot the case $\omega_B/\omega_A = 3/4$, in blue. We obtain a worse SNR but still can observe a violation of all the TURs derived for time-symmetric processes.

The only TURs that can set a bound that cannot be violated by $U_3$ are those obtained without posing the symmetry between the forward and backward process. In fact, the TURs in Equations~(\ref{landi}) and~(\ref{loose}) have been generalized, respectively, in Refs.~\cite{sam,franci}  by 
 releasing the assumption that forward and backward processes share the same distribution of the stochastic variables. These new bounds are given by
\begin{linenomath} 
    \begin{equation} \label{landigeneral}
        \frac{\text{var}(W)+\text{var}(W)_B}{(\langle W \rangle + \langle W \rangle_B)^2} \geq \frac{1}{2}\text{csch}^2[g(a/2)]
    \end{equation}
\end{linenomath}
and
\begin{linenomath} 
    \begin{equation} \label{loosegeneral}
        \frac{\text{var}(W)+\text{var}(W)_B}{(\langle W \rangle + \langle W \rangle_B)^2} \geq \frac{1}{e^{a/2}-1},
    \end{equation}
\end{linenomath}
where the quantities with subscript $B$ are referred to the backward process and $a = (\langle\Sigma\rangle + \langle\Sigma\rangle_B)/2$. In the case of $U_3$, the statistics of $W$ for the backward process are easily found since $U_3^{-1} = \tilde{U}_3$. Hence, $U_3^{-1}$ outputs the same work statistics as $U_3$ provided that systems $A$ and $B$ are swapped. Then, $\langle W_3 \rangle_B$, $\text{var}(W_3)_B$ and $\langle \Sigma_3 \rangle_B$ can be obtained from Equations~(\ref{superergo}),~(\ref{s3}) and~(\ref{fluct3}) simply swapping labels $A$ and $B$. Note also that the bounds (right-hand sides) given by the TURs in Equations~(\ref{landigeneral}) and~(\ref{loosegeneral}) depend only on the products $\beta\omega$, while the corresponding bounded quantities depend also on the frequency ratio $\omega_B/\omega_A$.

\begin{figure}[H]
    \includegraphics[width=10 cm]{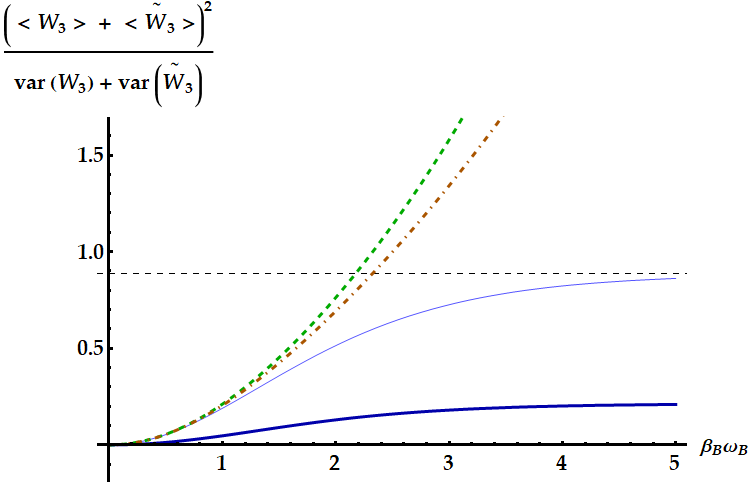}
    \caption{Ratio between the squared sum of the mean works extracted in the forward ($\langle W_3 \rangle$) and backward ($\langle \tilde{W}_3 \rangle$) processes and the sum of the corresponding variances as a function of $\beta_B\omega_B$ (solid lines). We set $\beta_A\omega_A = 10^{-3}$. We display the cases $\omega_B/\omega_A = 10^{-2}$ (dark-blue thick line) and $\omega_B/\omega_A = 10^{2}$ (light-blue thin line). The dot-dashed brown and the dashed green curve represent the upper bounds given by Equations~(\ref{landigeneral}) and (\ref{loosegeneral}), respectively. The dashed black horizontal line identifies the asymptotic value, which amounts to $8/9$ and is achieved for both $\beta_B\omega_B,$ and $\omega_B/\omega_A \rightarrow \infty$.}
    \label{franci}
\end{figure}

In Figure~\ref{franci}, we compare the reciprocal of the left-hand sides
of Equations~(\ref{landigeneral}) and~(\ref{loosegeneral}) for $U_3$ with
the corresponding bounds as a function of $\beta_B\omega_B$ with fixed
$\beta_A\omega_A \ll 1$. In this regime, $U_3$ maximizes the SNR. We show the two limiting cases $\omega_B/\omega_A \ll 1$ (thick dark-blue curve) and $\omega_B/\omega_A \gg 1$ (thin light-blue curve) together with the bounds obtained from the TURs in Equations~(\ref{landigeneral}) and~(\ref{loosegeneral}), identified by the dot-dashed brown curve and the dashed green curve, respectively. Note that these TURs are never violated and, as expected, the first is tighter than the second. Having set $\beta_A\omega_A \sim 0$, the maximum is asymptotically reached for $\beta_B\omega_B \gg 1$ and $\omega_B \gg \omega_A$, and amounts to $8/9$ (dashed horizontal line in Figure~\ref{franci}).

\section{Conclusions} \label{c}
We devised a consistent description of ergotropic heat engines for the
optimal work extraction from a couple of quantum systems, which are
cyclically restored to the canonical equilibrium at two different
temperatures. We provided an exhaustive study for the case of two
qutrits with equally-spaced energy levels by deriving the optimal
ergotropic transformations, the statistics of the extracted work and
the mean entropy production. 
We showed that going beyond the standard swap Otto
  engine allows one to improve the work extraction and also to enlarge
  the range of operation of the heat engine. We think that further interesting
  results for systems with arbitrary energy-level structures may be
  found by means of the approach outlined in Ref.~\cite{uz}.
Within the approach of stochastic
  thermodynamics we exploited a two-point measurement scheme to
  retrieve the first and second moment of the work distribution. We
  recall that, to this aim, many equivalent measurement schemes exist~\cite{allah3}. In future developments, it will be interesting to consider the effect of measurements explicitly performed on the
  quantum systems to monitor the engine, along with its impact on the
  thermodynamic cycles as performed, for example, in Ref.~\cite{son}.

We focused on the relative fluctuations of the work extracted by each
ergotropic transformation and showed that one of them, the double swap
$U_3$, violates many common TURs, specifically those based on the
assumption that the distributions of the extracted work for the
forward and backward processes are the same.

The application of our procedure to systems with higher dimensions is
promising because it will lead to the generalization of the ergotropic
transformations found for the qutrit case and will allow to finding new
transformations which, as shown in this work, may possibly extract
more work on average with lower fluctuations with respect to Otto
engines based on the swap interaction with qudits.





\vspace{6pt} 



\authorcontributions{Conceptualization, Massimiliano Federico Sacchi; methodology, Massimiliano Federico Sacchi; validation, Chiara Macchiavello; formal analysis, Giovanni Chesi; investigation, Giovanni Chesi and Massimiliano Federico Sacchi; writing---original draft preparation, Giovanni Chesi; writing---review and editing, Chiara Macchiavello and Massimiliano Federico Sacchi; visualization, Chiara Macchiavello and Massimiliano Federico Sacchi; supervision, Chiara Macchiavello and Massimiliano Federico Sacchi; project administration, Chiara Macchiavello and Massimiliano Federico Sacchi; funding acquisition, Chiara Macchiavello and Massimiliano Federico Sacchi. All authors have read and agreed to the published version of the manuscript.} 

\funding{This 
 research and the APC were funded by EU H2020 QuantERA ERA-NET Cofund in Quantum Technologies project QuICHE grant number 731473.} 

\dataavailability{Data sharing not applicable.
No new data were created or analyzed in this study. Data sharing is not applicable to this article.} 

\acknowledgments{This material is based upon work supported the Italian MUR through PRIN 2022. C.M. acknowledges support from the PNRR MUR Project PE0000023-NQSTI.}

\conflictsofinterest{The authors declare no conflict of interest. The funders had no role in the design of the study; in the collection, analyses, or interpretation of data; in the writing of the manuscript; or in the decision to publish the results.} 





\appendixtitles{no} 
\appendixstart
\appendix
\section[\appendixname~\thesection]{} \label{appA1}
Let us describe the two-stroke Otto engines. We consider two quantum systems $A$ and $B$ (with Hamiltonians $H_A$ and $H_B$) initially at thermal equilibrium with their own reservoirs $R_A$ and $R_B$ (with Hamiltonians $H_{R_A}$ and $H_{R_B}$) at inverse temperatures $\beta _A$ and $\beta _B$. Without loss of generality, we take $\beta _A < \beta _B$. We assume weak coupling between systems and reservoirs so that we can represent the initial state as the tensor product of canonical density matrices, namely
\begin{eqnarray}
 \rho_0 \otimes \rho_R  = \frac {1}{Z_A Z_B Z_{R_A} Z_{R_B}} e^{-\beta _A H_A} \otimes
e^{-\beta _B H_B} \otimes   e^{-\beta _A H_{R_A}} \otimes 
  e^{-\beta _B H_{R_B}}
\;,\label{tens}
\end{eqnarray}
where $Z_X = \text{Tr}[e^{-\beta_XH_X}]$.
We perform a two-stroke cyclic heat engine by $(i)$ isolating the two
quantum systems from the reservoir at $t=0^+$; $(ii)$ extracting work by a unitary transformation $U$ acting
on the two systems up to time $t=\tilde t$; $(iii)$ reconnecting the two quantum systems to
their respective reservoirs by weak coupling
to achieve complete thermalization at $t=t' \gg \tilde t$. We
remark that the unitary $U$ incorporates the free evolution of the two
systems and their interaction obtained by external (possibly
time-dependent) driving protocols, with the condition of being cyclic,
namely, such that initial and final Hamiltonian coincide,
i.e., $H_X=H_X(0)=H_X(\tilde t)$ for both $A$ and $B$.

The average value $\langle W \rangle $ of the work extracted on a
cycle corresponds to the opposite of the variation of the internal
energy of $A$ and $B$ during the unitary stroke $U$, i.e.,
\begin{eqnarray}
\langle W
\rangle = -\langle \Delta E_A \rangle -\langle \Delta E_B \rangle =
\Tr [(H_A +H_B)\rho_0 ] - \Tr [(H_A +H_B)U\rho_0 U^\dag ]
\;.\label{}
\end{eqnarray}
During the thermalization stroke, each system comes back to
equilibrium, namely system $A$ absorbs the average heat $\langle Q_H
\rangle = \Tr[H_A \rho _0]- \Tr [H_A U \rho _0 U^\dag ]= -\langle
\Delta E_A \rangle $ from the hot reservoir, and system $B$ dumps
$\langle Q_C \rangle =\Tr [H_B \rho _0]- \Tr [H_B U \rho _0 U^\dag
]=-\langle \Delta E_B \rangle $ on the cold one. In our convention
the cycle operates as a heat engine when $\langle W \rangle >0$,
$\langle Q_H \rangle >0$, and $\langle Q_C \rangle <0$. Clearly, the
first law is obtained as $\langle W\rangle = \langle Q_H \rangle +
\langle Q_C \rangle $. At each cycle, the two quantum systems come back
to their respective equilibrium states, and hence the average entropy
production per cycle simply corresponds to $\langle \Sigma \rangle =
-\beta_A \langle Q_H \rangle - \beta_B \langle Q_C \rangle = (\beta_A
- \beta_B) \langle \Delta E_A \rangle - \beta_B \langle W \rangle $.
 
Let us now describe the above thermodynamical cycle by a set of
stochastic trajectories which correctly reproduce the mean
values of all thermodynamic variables by an average of stochastic
variables over all possible trajectories.  We adopt an operational
approach based on complete energy measurements at different times
\cite{stoc3,th,camp} as in the typical derivation of Jarzynski
equality~\cite{j97}. This approach will allow us to study the complete
statistics of work extraction and heat exchanges, and
in particular to evaluate the fluctuations of work
and their relation with entropy production. 

\vspace{-5pt}
\begin{figure}[H]
    \includegraphics[width=12.5 cm]{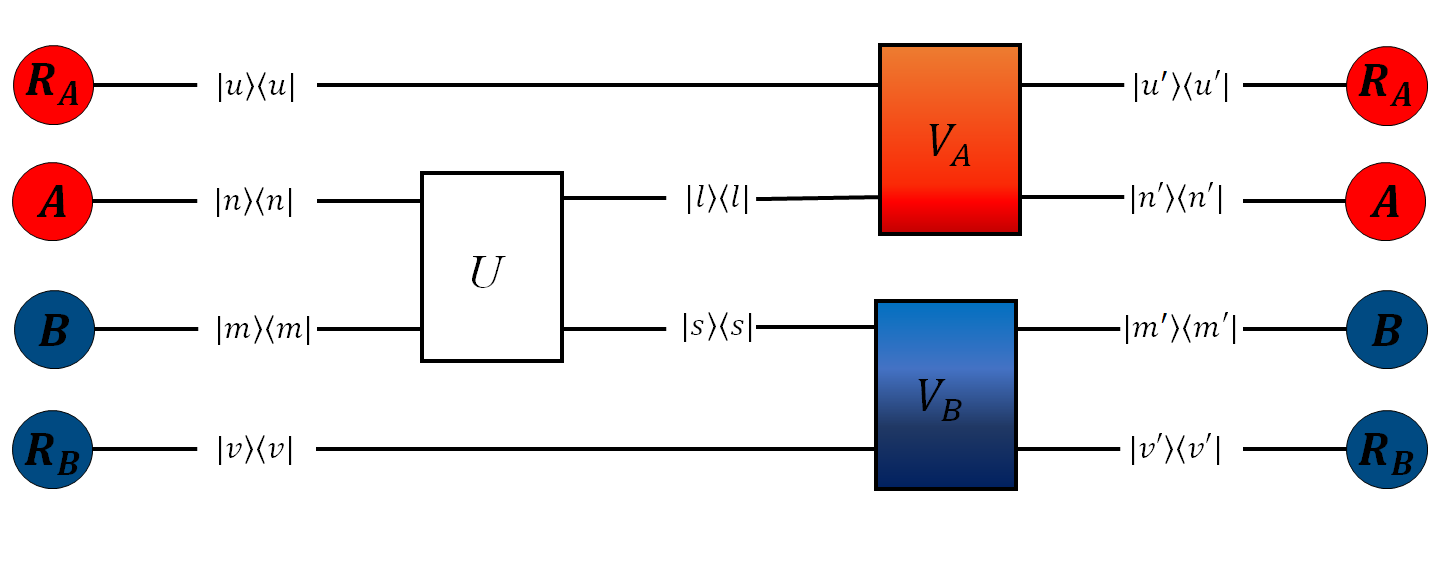}
    \caption{Scheme for the stochastic description of a cycle of the Otto two-stroke engine. Labels $A$ and $B$ identify the two systems operated by the engine as a working fluid, $R_A$ and $R_B$ are the corresponding reservoirs. The unitary $U$ is the transformation extracting work, while $V_A$ and $V_B$ are the energy-preserving unitaries for the thermal relaxation of each system with its pertaining reservoir.}
    \label{eng}
\end{figure}

As depicted in Figure~\ref{eng}, we identify a single stochastic trajectory by
the outcomes of the sequential fine-grained energy measurements of 
$H_A,H_B,H_{R_A},H_{R_B}$ at the beginning of the cycle $t=0^+$; of 
$H_A,H_B$ at the end of the work stroke $t=\tilde t$ operated by $U$;  
and finally of $H_A,H_B,H_{R_A},H_{R_B}$ at the end of the
thermalization stroke $t=t'$. We denote by $V_X$, with $X=A,B$, the
unitary operator representing the joint evolution of system $X$ and
reservoir $R_X$ by weak coupling and energy-preserving interaction 
up to complete thermal equilibrium at $t=t'\gg \tilde t$. By respecting the order of the above measurements, we denote by $\gamma $ the
 stochastic trajectory corresponding to the sequence of outcomes, namely
\begin{eqnarray}
  \gamma =\{n,m,u,v;l,s;n',m',u',v'\}
\end{eqnarray}
with corresponding energy eigenvalues
\begin{eqnarray}
\{E_n^A,E_m^B,E_u^{R_A},E_v^{R_B};
  E_l^A,E_s^B; E_{n'}^A,E_{m'}^A,E_{u'}^{R_A},E_{v'}^{R_B}\}
\;.\label{}
\end{eqnarray}
We could denote by $|i \rangle _C$ the eigenvector pertaining to the
eigenvalue $E_i^C$ of Hamiltonian $H_C$, with $C=A,B,R_A,R_B$, but
we will generally write $|i \rangle $ instead of $|i \rangle _C$ since in the
following it will be clear from the context (and with the help of
Figure~\ref{eng}) the respective Hilbert space of all eigenvectors. 

Let us now evaluate the probability $P[\gamma ]$ of occurrence of
a specific trajectory $\gamma $. Since the initial state is given by Equation~(\ref{tens}),
the probability $p(n,m,u,v)$ for the initial
outcomes $n,m,u,v$ is given by the product of Gibbs weights, namely
\begin{eqnarray}
  p(n,m,u,v)=\frac{1}{Z_A Z_B Z_{R_A} Z_{R_B}}
  e^{-\beta _A (E^A_n +E_u^{R_A})}\,e^{-\beta _B (E^B_m +E_v^{R_B})}\;.
\end{eqnarray}
The conditional probability $q_w(l,s|n,m)$ pertaining to the energy
measurements of $A$ and $B$ with outcomes $l$ and $s$ after the unitary
stroke $U$, given initial outcomes $n$ and $m$, writes 
\begin{eqnarray}
  q_w(l,s|n,m)
  = |\langle l | \langle s| U |n \rangle |m \rangle |^2
  \;.\label{qw}
\end{eqnarray}
Finally, the conditional probability for the thermalization
stage is given by
\begin{eqnarray}
  q_t(n',m',u',v'|l,s,u,v)=
|\langle n' | \langle u' | V_A | l \rangle
 | u \rangle |^2  \,  |\langle m' | \langle v' | V_B | s \rangle
 | v \rangle |^2 \;.\label{qtt}
\end{eqnarray}
It follows that the probability of the trajectory
$\gamma =\{n,m,u,v;l,s;n',m',u',v'\}$ is given by
\begin{eqnarray}
  P[\gamma ]=  p(n,m,u,v)  q_w(l,s|n,m)  q_t(n',m',u',v'|l,s,u,v)
  \;.\label{pgg}
\end{eqnarray}
One easily identifies the functions of stochastic variables in
correspondence to the thermodynamical variables of interest for each
trajectory. 
Clearly, the work contribution corresponds~to 
\begin{equation}
W[\gamma]=E^A_n-E^A_l+E^B_m-E^B_s. 
\end{equation}
On the other hand,
the heat released by reservoirs $A$ and $B$ corresponds to
\begin{eqnarray}
&&Q_H[\gamma
  ]=E^{R_A}_u -E^{R_A}_{u'}\simeq E^A_{n'}-E^A_l \;,\nonumber \\& &
Q_C[\gamma
]=E^{R_B}_v -E^{R_B}_{v'} \simeq E^B_{m'}-E^B_s
\;,\label{simeq}
\end{eqnarray}
since we describe the thermalization with each reservoir by weak
coupling and energy-preserving interactions
\cite{jar2}. Notice that under this approximation no
  work is generated by connecting and disconnecting the systems with
  the reservoirs~\cite{jukka,gabri,moli}.

By weighting each possible trajectory with its probability of occurrence
we obtain the joint probability for extracting work $W$ along with heat
exchanges $Q_H$ and $Q_C$ as follows
\begin{eqnarray}
&& p(W,Q_H,Q_C)= \sum _{\gamma  }P[\gamma ] \delta (W -
  (E^A_n-E^A_l)-(E^B_m-E^B_s))\nonumber \\& & \times \, 
\delta (Q_H -(E^A_{n'}-E^A_l))
\delta(Q_C -(E^B_{m'}-E^B_s))
\;.\label{p3}
\end{eqnarray}
Since $V_X$ models the complete thermalization by the reservoir $R_X$, one has
\begin{eqnarray}
\Tr _{R_X}\left[V_X \left (\sigma \otimes \frac {1}{Z_{R_X}}e^{-\beta
    _X H_{R_X}} \right ) V_X^\dag\right ] =
\frac {1}{Z_{X}}e^{-\beta _X H_{X}}
\;\label{}
\end{eqnarray}
for arbitrary density matrix $\sigma $ of system $X$. This fact can be used
to simplify Equation~(\ref{p3}) by summing on all reservoir indexes
$\{u,u',v,v'\}$. Hence, $p(W,Q_H,Q_C)$ can be rewritten in terms of
measurements outcomes only on systems $A$ and $B$, namely 
\begingroup\makeatletter\def\f@size{9}\check@mathfonts
\def\maketag@@@#1{\hbox{\m@th\normalsize\normalfont#1}}%
\begin{eqnarray}
&&  p(W,Q_H,Q_C)
  = \sum _{n,m,l,s,n',m'}
\frac{1}{Z_A ^2 Z_B ^2}
e^{-\beta _A (E^A_n +E_{n'}^A)
  -\beta _B (E^B_m + E_{m'}^B)}\;
|\langle l | \langle s| U |n \rangle |m \rangle |^2\,\times 
\nonumber \\& &   \delta (W -
(E^A_n-E^A_l)-(E^B_m-E^B_s))\,
\delta (Q_H -(E^A_{n'}-E^A_l))\,
\delta(Q_C -(E^B_{m'}-E^B_s))
\;.\label{pwcc}
\end{eqnarray} 
\endgroup
Equation~(\ref{pwcc}) allows one to study the complete statistics of an
ergotropic heat engine. The first principle of thermodynamics for the
cycle is recovered since the average variation of the internal energy
$\langle \Delta U \rangle = \langle Q_H +Q_C -W \rangle $ correctly
gives zero, as shown as follows
\begingroup\makeatletter\def\f@size{9}\check@mathfonts
\def\maketag@@@#1{\hbox{\m@th\normalsize\normalfont#1}}%
\begin{eqnarray}
&&  \langle \Delta U \rangle =
  \langle Q_H +Q_C -W \rangle=
  \int d W \int d Q_H \int dQ_C \,p(W,Q_H,Q_C) \,(Q_H +Q_C -W)
  \nonumber \\& & 
= \sum_{n,n',m,m'} \frac{1}{Z_A ^2 Z_B ^2}
e^{-\beta _A (E^A_n +E_{n'}^A)
  -\beta _B (E^B_m + E_{m'}^B)} (E^A_{n'}-E^A_n +E^B_{m'}-E^B_m)
=0  
\;.\label{}
\end{eqnarray}
\endgroup
One also has $\langle Q_H \rangle = -\langle \Delta E_A \rangle $,
where $\langle \Delta E_A \rangle $ is the average variation of the
internal energy of $A$ during the unitary stroke, whose expectation can
be obtained by averaging $E_l^A-E^A_n$ over all
trajectories. Similarly, $\langle Q_C \rangle = -\langle \Delta E_B
\rangle $, where $\langle \Delta E_B \rangle $ has corresponding
stochastic values given by $E_s^B-E^B_m$.

By introducing the trajectories for the thermalization stroke, we remark that the present result allows us to refine the approach of Refs.~\cite{landi,max,max2}, where the stochastic values of $Q_H$ were identified with $-\Delta E_A$ (and analogously for $Q_C$ with $-\Delta E_B$). In fact, notice that the relation $W=Q_H +Q_C$ does not generally hold at the trajectory level. Anyway, since $\langle Q_H + \Delta E_A\rangle $ corresponds to the average of $E^A_{n'}-E_n^A$ over the trajectories, for increasing number of cycles the discrepancy between the stochastic variables $Q_H $ and $-\Delta E_A$ remains bounded by the finite energy of system $A$, whereas both $Q_H $ and $\Delta E_A$ increase linearly with the number of cycles, thus providing $\langle Q_H \rangle + \langle \Delta E_A \rangle =0$. Analogous point applies for the variables $ Q_C $ and  $\Delta E_B $.

All results about the stochastic efficiency $\eta$ of the heat engines given in Refs.~\cite{max,max2} rigorously hold for $\eta $ defined in terms of $\Delta E_A$, namely $\eta = - W/\Delta E_A$. One could refine and compare the results for $\eta
$ defined as $\eta = W/Q_H$ by means of the probability distribution presented here in Equation~(\ref{pwcc}). The subtle difference between these two definitions of stochastic efficiency was already discussed in Ref.~\cite{jukka}, where the thermalization in a two-stroke Otto engine was modeled by the quantum jump method.

The probability of work extraction is the marginal of $p(W,Q_H,Q_C)$
in Equation~(\ref{pwcc}) with respect to the heat exchanges, and one has 
\begin{eqnarray}
&&  p(W)=\int d Q_H \int d Q_C \,p(W,Q_H,Q_C)=
\nonumber \\& &   \sum _{n,m,l,s}
\frac{1}{Z_A Z_B }
e^{-\beta _A E^A_n 
  -\beta _B E^B_m }\;
|\langle l | \langle s| U |n \rangle |m \rangle |^2\,  \delta (W -
(E^A_n-E^A_l)-(E^B_m-E^B_s))\;. \quad \quad \label{pw}
\end{eqnarray}
Let us now consider a backward protocol where the measurements on the
quantum systems and the reservoirs are performed in the reverse
ordering, along with the time-reversal evolution of all interactions.
The initial state for the backward protocol is again taken as the
product of canonical density matrices, namely as in
Equation~(\ref{tens}). By assuming that all Hamiltonians are invariant
under time-reversal at all times~\cite{andrie,camp,stoc2}, the backward
protocol is then equivalent to follow Figure~\ref{eng} from the right to the
left, along with the replacement of $V_A$, $V_B$ and $U$ with
$V_A^\dag$, $V_B^\dag $ and $U^\dag $, respectively.

\par We can compare the forward and the backward protocols by the
probability $P[\gamma ]$ of a trajectory $\gamma
=\{n,m,u,v;l,s;n',m',u',v'\}$ and the probability $P_B [\gamma _B]$
for the occurrence of the specular reverse trajectory with the same
measurement outcomes, namely $\gamma _B =\{n',m',u',v';l,s;n,m,u,v\}$.
The comparison between the forward and the backward protocol is made
by the logarithm of the probabilities, which defines the stochastic
entropy $\Sigma [\gamma]$ generated along a trajectory $\gamma$ as~\cite{stoc2}
\begin{eqnarray}
\Sigma [\gamma ]= \log \frac{P[\gamma ]}{P_B[\gamma _B]}\;.\label{defs}
\end{eqnarray}
Since each $\gamma $ identifies a corresponding backward trajectory
$\gamma _B$, the fluctuation theorem simply follows as
\vspace{6pt}
\begin{eqnarray}
  \langle e^{-\Sigma} \rangle = \sum  _{\gamma _B} P_B[\gamma _B]=1\;.
\label{intf}  
\end{eqnarray}
In the present case, using Equation~(\ref{pgg}) and the cancellation between
forward and backward conditional probabilities for both $q_w$ and
$q_t$, one obtains
\begin{eqnarray}
  \frac{P[\gamma ]}{P_B[\gamma _B]}= \frac {p(n,m,u,v)}{p(n',m',u',v')}
  \;,\label{}
\end{eqnarray}
and hence
\begin{eqnarray}
&&  \Sigma [\gamma ] = \beta _A (E^A_{n'}-  E_{n}^A +E^{R_A}_{u'} - E^{R_A}_{u}) 
  + \beta _B (E^B_{m'}-  E_{m}^B +E^{R_B}_{v'} - E^{R_B}_{v})
  \nonumber \\& & \simeq    \beta _A (E^A_{l}-  E_{n}^A) 
  + \beta _B (E^B_{s}-  E_{m}^B )  
  \;, \label{}
\end{eqnarray}
where we used Equation (\ref{simeq}).  
As for the case of the stochastic work $W[\gamma ]=E^A_{n}- E_{l}^A +
E^B_{m}- E_{s}^B $, notice that also $\Sigma [\gamma ]$ depends only on the
reduced set of indexes $\{n,m,l,s \}$ pertaining to the unitary
stroke operated by $U$. The probability of the stochastic entropy is
then given by
\begingroup\makeatletter\def\f@size{9}\check@mathfonts
\def\maketag@@@#1{\hbox{\m@th\normalsize\normalfont#1}}%
\begin{eqnarray}
  p(\Sigma )=
  \sum _{n,m,l,s}
\frac{1}{Z_A Z_B }
e^{-\beta _A E^A_n 
  -\beta _B E^B_m }\;
|\langle l | \langle s| U |n \rangle |m \rangle |^2 \,
\delta (\Sigma - \beta _A (E^A_n-E^A_l)- \beta _B (E^B_m-E^B_s))  
  \;. \quad \label{pss}
\end{eqnarray}
\endgroup
From Equation~(\ref{pss}), one correctly recovers the equivalent identities $\langle \Sigma
\rangle = \beta_A \langle \Delta E_A \rangle + \beta _B \langle \Delta
E_B \rangle = (\beta_A - \beta_B) \langle \Delta E_A \rangle - \beta_B \langle
W \rangle = -\beta _A \langle Q_H \rangle - \beta_B \langle Q_C
\rangle $. 

Since from definition (\ref{defs}) one has $\Sigma [\gamma _B]=-\Sigma [\gamma
]$ we easily derive the detailed fluctuation theorem as follows  
\begin{eqnarray}
&& p(\Sigma )=\sum _{\gamma } P[\gamma ] \delta (\Sigma -\Sigma [\gamma
])=\sum _{\gamma }e^{\Sigma [\gamma]} P_B[\gamma _B] 
\delta (\Sigma -\Sigma [\gamma ])\nonumber \\& &
=e^{\Sigma }\sum _{\gamma }
P_B[\gamma _B] 
\delta (\Sigma  + \Sigma [\gamma _B])=e^{\Sigma }p_B (- \Sigma )
\;.\label{}
\end{eqnarray}
Indeed, by analogous derivation, for any set $\{X_i [\gamma ] \}$ of
odd stochastic variables such that $X_i[\gamma _B]=-X_i[\gamma
]$, one has 
\begin{eqnarray}
p(\{ X_i\},\Sigma)=e^{\Sigma } p_{B}(\{-X_i\}, -\Sigma)\;.\label{exch}
\end{eqnarray}
In particular, since $\Sigma = (\beta _A -\beta _B)\Delta E_A - \beta
_B W$, we can also write~\cite{andrie,cth,sini,frq}
\begin{eqnarray}
\frac {p(W,\Delta E_A)}{p_B(-W,- \Delta E_A)}= e^{\Sigma }  \;.\label{sini}
\end{eqnarray}
where
\begin{eqnarray}
&& p(W,\Delta E_A)= \sum _{n,m,l,s} \frac{1}{Z_A Z_B } e^{-\beta _A
    E^A_n -\beta _B E^B_m }\; |\langle l | \langle s| U |n \rangle |m
  \rangle |^2\, \nonumber \\& & \times\,\delta (W -
  (E^A_n-E^A_l)-(E^B_m-E^B_s))\, \delta (\Delta E_A - (E^A_l-E^A_n))
  \;\label{pjoint}
\end{eqnarray}
and $p_B$ identifies the distribution of the backward process, described by the transformation $U^{\dagger}$ instead of $U$.
We recall here that the relation $\Delta E_B = -W -\Delta E_A$ holds for
the stochastic variables, namely at the trajectory level.
 
When the unitary operator $U$ is of the form 
\begin{eqnarray}
U=e^{i\phi _A H_A + i \phi_B H_B} V e^{i\psi _A H_A + i \psi_B H_B}
\;,\label{}
\end{eqnarray}
with unitary Hermitian $V=V^\dag $ and arbitrary phases $\phi_A,\phi
_B,\psi_A,\psi_B$, notice that one has the~symmetry
\vspace{6pt}
\begin{eqnarray}
  p_B(W,\Delta E_A)=p(W,\Delta E_A)
  \;.\label{bf}
\end{eqnarray}
Typically, this happens when the
time-dependent protocol that actualizes the unitary evolution $U$ is a
time-symmetric driving~\cite{cth,andrie,stoc3}. We remark that the
TUR in Equation~(\ref{landi}) derived from the fluctuation
theorem of Equation~(\ref{exch}) implicitly assumed the condition $p_B
(\{ X_i\},\Sigma)=p(\{X_i\}, \Sigma)$.

Let us now consider in more detail the joint probability $p(W,\Delta
E_A)$. The full statistics of $W$ and $\Delta E_A$ is equivalently
contained in the characteristic function $\chi(\lambda ,\mu)$ given by the
Fourier transform 
\begin{eqnarray}
\chi(\lambda , \mu )=\int dW \int d\Delta E_A \,
p(W,\Delta E_A)\, e^{i \lambda W + i \mu \Delta E_A} \;. \label{tf}
\end{eqnarray}
Here, $\lambda$ and $\mu$ denote the
counting parameters for $W$ and $\Delta E_A$, so that all moments and
correlations can be recovered as
\begin{eqnarray}
  \langle W^j \Delta E_H ^k \rangle = (-i)^{j+k}\left.
\frac{\partial  ^{j+k} \chi (\lambda ,\mu )}{\partial \lambda ^j
    \partial \mu ^k}\right
  |_{\lambda=\mu =0}
\label{mom}\;.
\end{eqnarray}
Using Equation~(\ref{pjoint}) and applying the delta functions in the integrals of Equation~(\ref{tf})
one obtains 
\begin{eqnarray}
  \chi(\lambda , \mu )&&=\frac{1}{Z_A Z_B}
  \sum _{n,m,l,s}   e^{-\beta _A E_n^A}e^{-\beta _B E_m^B} 
  e^{i \lambda (E_n^A -E_l^A
    +E_m^B-E_s^B)} e^{i\mu (E_l^A- E_n^A)}
      \nonumber \\& & \times 
  \Tr [U^\dag  (|l \rangle \langle l|
    \otimes |s \rangle \langle s|) U  (|n \rangle \langle
    n| \otimes |m \rangle \langle m|)] \nonumber \\& &=
\Tr [U^\dag   (e^{-i (\lambda -\mu) H_A}
  \otimes e^{- i \lambda H_B})
    U  (e^{i (\lambda - \mu )H_A}\otimes e^{i \lambda H_B})\,\rho_0]
      \;.
\label{a5}
\end{eqnarray}
Recalling Equation~(\ref{tens}), one easily verifies the identity $\chi
[-i\beta _B , i(\beta _A -\beta _B) ]=1$, which corresponds to the
fluctuation theorem of Equation~(\ref{intf}). In fact, one has
\begin{eqnarray}
\langle e^{-\Sigma
}\rangle = \int dW \int \Delta E_A \,
p(W,\Delta E_A)\, e^{  \beta_B W -(\beta_A - \beta_B)\Delta E_A 
}= \chi [-i\beta _B , i(\beta _A -\beta
  _B) ]=1\;.
\end{eqnarray}
In terms of the characteristic function, we notice that the detailed
fluctuation theorem of Equation~(\ref{sini}) is translated into the symmetry 
\begin{eqnarray}
\chi _R ( \lambda , \mu )=\chi [-i\beta _B - \lambda , i(\beta _A -\beta _B)
  -\mu ] \;.\label{}
\end{eqnarray}

In the presence of a symmetry in the unitary stroke achieved by $U$
such that for a real $x \neq 0$ one has
\begin{eqnarray}
  [H_A + x H_B,U ] = 0
\;,\label{sima}
\end{eqnarray}
from Equation~(\ref{a5}) one obtains the corresponding property
\begin{eqnarray}
  \chi (\lambda , \mu )=  \chi ((1-x)\lambda + x \mu ,(1-x)\lambda + x \mu )
=  \chi (0,\mu -(1-x^{-1})\lambda )
\;,\label{idet}
\end{eqnarray}
i.e., the characteristic function becomes a function of a single
variable. It follows that $x \partial _\lambda \chi = (1-x) \partial
_\mu \chi $, and from Equation~(\ref{mom}) one obtains the symmetry
relations
\begin{eqnarray}
  \langle W^j \Delta E_A^k \rangle
  =\left ( \frac {x}{1-x }\right ) ^{k} \langle W^{j+k} \rangle =
\left ( \frac {1-x}{x }\right ) ^{j} \langle \Delta E_A ^{j+k} \rangle 
\;,\label{sym}
\end{eqnarray}
namely the stochastic variables $W$ and $\Delta E_A$ are perfectly
correlated. Moreover, since $\Delta E_B =-W-\Delta E_A$, one has
$\langle \Delta E_A ^{k} \rangle = (-x)^k \langle \Delta E_B ^{k}
\rangle $. It also follows that the average entropy is simply
proportional to the average work, namely
\vspace{6pt}
\begin{eqnarray}
  \langle \Sigma \rangle = \frac{x \beta _A -\beta _B}{1-x}  \langle W \rangle \;.\label{entp} 
\end{eqnarray}
The effect of a strong symmetry as Equation~(\ref{sima}) can be seen
directly on the joint probability $p(W,\Delta E_A)$. 
In fact, by using the last expression in Equation~(\ref{idet}) for
$\chi(\lambda ,\mu)$ in the
inverse Fourier transform of Equation~(\ref{tf}), one easily obtains
\begin{eqnarray}
p(W,\Delta E_A)=p(\Delta E_A)\, \delta (W+(1-x^{-1})\Delta E_A)
\;,\label{core}
\end{eqnarray}
namely one has the perfect correlation $p(W| \Delta E_A)=\delta
(W+(1-x^{-1})\Delta E_A)$.  In this case, one also recognizes that the
stochastic $\Delta E_A$-efficiency defined as the ratio $\eta _{\Delta
  E_A}=\frac{W}{-\Delta E_A}$, is a self-averaging quantity and has no
fluctuations, since $\eta _{\Delta E_A}=1-x^{-1}$. Interestingly,
under full correlation between $W$ and $\Delta E_A$ (and hence between
$W$ and $\Sigma $), in Ref.~\cite{campisi21} it is shown that a general
lower bound for the mean entropy $\langle \Sigma \rangle $ in terms of
the asymmetry of the marginal work distribution $p(W)$ evaluated by the
relative entropy $D(p(W)\|p(-W))$ is saturated.



\begin{adjustwidth}{-\extralength}{0cm}

\reftitle{References}

\PublishersNote{}
\end{adjustwidth}
\end{document}